\documentclass[preprint,aps,floatfix,unsortedaddress]{revtex4}

\usepackage{verbatim}
\usepackage{graphicx}
\usepackage{wrapfig}
\usepackage{gensymb}
\usepackage{array}
\usepackage{color}

\usepackage{amsmath}
\usepackage{multirow}
\usepackage{amstext}

\usepackage[utf8]{inputenc}
\usepackage[T2A]{fontenc}
\usepackage{morefloats}

\newcommand{\cm}{cm$^{-1}$}
\newcommand{\n}{$\nu$}
\newcommand{\1}{$_{1}$}
\newcommand{\2}{$_{2}$}
\newcommand{\3}{$_{3}$}
\newcommand{\4}{$_{4}$}
\newcommand{\5}{$_{5}$}
\newcommand{\6}{$_{6}$}
\newcommand{\7}{$_{7}$}
\newcommand{\8}{$_{8}$}
\newcommand{\9}{$_{9}$}
\newcommand{\p}{^\prime}
\newcommand{\pp}{^{\prime\prime}}

\begin{document}

\title{A hybrid variational-perturbation calculation of the ro-vibrational spectrum
of nitric acid}
\author {A.I. Pavlyuchko$^{a,b}$, S.N. Yurchenko$^a$, Jonathan Tennyson$^a$ \\
$^a$ Department of Physics and Astronomy, University College London, London, WC1E 6BT, UK;\\
$^b$ Department of Physics, Moscow State University of Civil Engineering (MGSU), Russia, (pavlyuchko@rambler.ru)}

\date{\today}

\begin{abstract}

  Rotation-vibration spectra of the nitric acid molecule, HNO\3, are
  calculated for wavenumbers up to 7000~\cm. Calculations are
  performed using a Hamiltonian expressed in internal curvilinear
  vibrational coordinates solved using a hybrid variational-perturbation
  method. An initial potential energy surface (PES) and dipole moment
  function (DMF) are calculated {\it ab initio} at the
  CCSD(T)/aug-cc-pVQZ level of theory.  Parameters of the PES and DMF
  are varied to minimize differences between the calculated and
  experimental transition frequencies and intensities.  The average,
  absolute deviation between calculated and experimental values is
  0.2~\cm\ for frequencies in the fundamental bands and 0.4~\cm\ for
  those in the first overtone and lowest combination bands.  For the
  intensities, the calculated and experimental values differ by 0.3\%
  and 40\% for the fundamentals and overtones, respectively.  The
  optimized PES and DMF are used to calculate the room-temperature ro-vibrational
  spectrum.  These calculation reproduce both the form of the
  absorption bands, and fine details of the observed spectra, including the
  rotational structure of the vibrational bands and the numerous hot
  absorption band. Many of these hot bands
  are found to be missing from the compilation in HITRAN. A room temperature
  line list comprising $2 \times 10^{9}$ lines is computed.

\end{abstract}

\maketitle

\section{Introduction}
\label{s:intro}

Nitric acid (HNO\3), in spite of its low concentration,  makes a significant
contribution to the infrared (IR) spectrum of the Earth's atmosphere, since
it has a number of strong  absorption bands absorption lying in the water
transparency window
\cite{81Laxxxx.HNO3,
92GoMuBl.HNO3,
98GoRiPe.HNO3,
03DoIoNo.HNO3,
06FlBrCa.HNO3,
11WaBiHy.HNO3}.
Yet its spectrum remains poorly characterized with, for example, no transition
wavenumbers above 2000 \cm\ included in the HITRAN database \cite{jt557}, despite the
fact that several fundamental bands lie at higher wavenumbers. This issue is not due to any lack
of attempts to measure the IR spectrum of HNO\3
\cite{65CoRixx.HNO3,
78ChGixx.HNO3,
81GhBlBa.HNO3,
82VaPiHe.HNO3,
84MaWexx.HNO3,
85WeMaGu.HNO3,
88GiHaSc.HNO3,88BoCrDei.HNO3,88BoCrDe.HNO3,88GoBuHo.HNO3,88CrBoDe.HNO3,
89PeLaVa.HNO3,89Maxxxx.HNO3,
91TaLoLui.HNO3,91TaLoLu.HNO3,
92TaLoLu.HNO3,92MaWexx.HNO3,
93MaTaLo.HNO3,93PeJaVa.HNO3,
94TaLoLu.HNO3,94PeFlCA.HNO3,94CoElAt.HNO3,
95CoPexx.HNO3,
96TaWaLo.HNO3,96LoLaWa.HNO3,96GoOePe.HNO3,96PaCoGo.HNO3,
97WaOnCh.HNO3,97WaOnTa.HNO3,
98KePeFl.HNO3,98Pexxxx.HNO3,
99PeFlKe.HNO3,99LuPexx.HNO3,
01PeGoHe.HNO3,
03PeHeBu.HNO3,
04FeHaVa.HNO3,04PeOrFl.HNO3,
05PeHeBe.HNO3,
06PeMbxx.HNO3,06KoEuLe.HNO3,
08PeKiJo.HNO3,
09GoTrPe.HNO3,
10PeHeMe.HNO3,
13Pexxxx.HNO3},
but rather to do with the difficulty of interpreting its spectrum and making line assignments.
Experimental line intensities have also been the subject of number of studies
\cite{84GiVaGo.HNO3,89PeLaVa.HNO3,96WaLoTa.HNO3,98DoOrAm.HNO3,99PeFlKe.HNO3,03ChShBl.HNO3,03ToBrCo.HNO3,04PeOrFl.HNO3,06PeMbxx.HNO3}.

The study of the HNO\3 spectrum over a range of temperatures is a
difficult experimental and theoretical problem.
Experimental study of the HNO\3 IR spectrum is challenging because
in the gas phase it is a mixture containing significant numbers of
dimers and complexes, as well as the products of its dissociation
(NO\2, H\2O, O\2).  Therefore, experimental HNO\3 spectra  are
usually processed spectra in which spectra due to dimers, complexes
and dissociation products have been subtracted.  In addition, HNO\3 is
a chemically aggressive species, which greatly complicates the
experimental study of its spectrum at higher temperatures.

From the theoretical perspective, study of the ro-vibrational infrared
spectrum of the HNO\3 is difficult because of the relatively large number
of the vibrational degrees of freedom, $N_c=9$, and the large
anharmonicity of its vibrations.  There are only limited attempts to
solve the vibrational and ro-vibrational problems using
full-dimensionality
\cite{99BeVexx.HNO3,06KoEuLe.HNO3,10LaNaxx.HNO3,11AvCaxx.HNO3,12AvCaxx.HNO3}.
Benderskii and Vetoshkin \cite{99BeVexx.HNO3} used a perturbative
approach to study the tunneling dynamics of internal rotation.
Lauvergnat and Nauts \cite{10LaNaxx.HNO3} also concentrated
on these levels in both reduced and full dimensionality. Konen {\it et al}
\cite{06KoEuLe.HNO3} used second-order vibrational
perturbation theory (VPT2) to help interpret their experimental
findings. More recently, Avila and Carrington
\cite{11AvCaxx.HNO3,12AvCaxx.HNO3} have performed full-dimensional
variational calculations with a particular focus on how to make such
studies efficient. None of these studies considered transition
intensities and amongst various {\it ab initio} studies using more
approximate treatments \cite{92LeRixx.HNO3, 95GrLeHe.HNO3,
05MiChGe.HNO3, 07BiWaHy.HNO3, 09GuDeRu.HNO3, 10NoSuHu.HNO3}, only
Lee and Rice \cite{92LeRixx.HNO3} appear to have considered (harmonic)
intensities.

Recently, we \cite{jt588} developed a hybrid
variational-perturbational calculation scheme for computing IR spectra
of polyatomic species.  HNO\3 was one of the species used to test this
methodology. Here we present calculations of the infrared spectrum of
HNO\3 performed using this method.  The calculations provide a
comprehensive room temperature line list covering the range 0 - 7000 \cm.

\section{Hamiltonian}
\label{s:hamiltonian}

Our vibration-rotation Hamiltonian written in curvilinear internal coordinates and an
Eckart embedding has the form \cite{88GrPa.method}
\begin{equation}
\hat H_{vr} = \hat H_v
- \frac{\hbar^2}{2} \sum_{a, b} \frac{\partial}{\partial \xi_a}
\mu_{ab} (\underline{q}) \frac{\partial}{\partial \xi_b}
~, ~ ~ ~ ~ ~ ~ \xi_a, \xi_b = \alpha, \beta, \gamma ~,
\end{equation}
where $\underline{\xi}$ are the rotational coordinates and
$\hat H_v$ is the vibrational part of the Hamiltonian
\begin{equation}
\label{e:fham}
\hat H_v = \hat T_v + V (\underline{q})
\end{equation}
\begin{equation}
\hat T_v =
- \frac{\hbar^2}{2} \sum_{i,j} t^{\frac{1}{4}} \frac{\partial}{\partial q_i}
                  g_{ij}(\underline{q}) t^{-\frac{1}{2}} \frac{\partial}{\partial q_j} t^{\frac{1}{4}}.
\end{equation}
Here $q_i$ are internal, vibrational curvilinear coordinates given by changes
in the bond lengths, valence bond angles, and dihedral angles  from the
corresponding equilibrium values; $\alpha$, $\beta$, $\gamma$ are the Euler
angles between the axes of the equilibrium moment of inertia tensor and
external Cartesian coordinate axes; $\mu_{ab}(\underline{q})$ are elements of
the inverse of the moment of inertia tensor, $\underline{I}(\underline{q})$;
$\hat T_v$ is the vibrational kinetic energy operator and
$g_{ij}(\underline{q})$ are elements of the kinetic energy coefficients matrix
$\underline{G}(\underline{q})$ and $t = \det[\underline{G}]$. Finally,
$V(\underline{q})$ is the molecular potential energy.

After transformation, the vibrational kinetic operator can be written as
\begin{equation}
\hat T_v =
- \frac{\hbar^2}{2} \sum_{i,j} \frac{\partial}{\partial q_i}
                  g_{ij}(\underline{q}) \frac{\partial}{\partial q_j}
+ \beta(\underline{q}),
\end{equation}
where
\begin{equation}
\label{e:pseodopot}
\begin{split}
{\beta(\underline{q})}&= -\frac{\hbar^2}{2} \sum_{i,j}{\left\{\frac{{\partial}{g}_{ij}(\underline{q})}{{\partial}{q_i}}
\sum_{k,l}{\zeta}_{kl}(\underline{q})\frac{{\partial}{g}_{kl}(\underline{q})}{{\partial}{q_j}}
\right.}+\\
+\frac{1}{4}{g}_{ij}(\underline{q})&\Bigl[\sum_{k,l}{\zeta}_{kl}(\underline{q})
\frac{{{\partial}^2}{g}_{kl}(\underline{q})}{{\partial}{q_i}{\partial}{q_j}}-
\sum_{k,l,m,n}{\zeta}_{kl}(\underline{q}){\zeta}_{mn}(\underline{q})\times\Bigr.\\
&\times\left.\Bigl.\Bigl(\frac{{\partial}{g}_{lm}(\underline{q})}{{\partial}{q_i}}
\frac{{\partial}{{g}_{kn}(\underline{q})}}{{\partial}{q_j}}+
\frac{{\partial}{{g}_{kl}(\underline{q})}}{{\partial}{q_i}}\frac{{\partial}
{g}_{mn}(\underline{q})}{{\partial}{q_j}}\Bigr)\Bigr]\right\},
\end{split}
\end{equation}
is the pseudo-potential or Watson term \cite{watsonterm}; $\zeta_{ij}(\underline{q})$  are
elements of $\underline{G}(\underline{q})^{-1}$.

We have performed calculations \cite{ap27, 88GrPa.method} which suggest
that the pseudo-potential Eq.~(\ref{e:pseodopot}) makes only a small
contribution to the vibrational energy levels of polyatomic molecules such as
HNO\3. For the water molecule this contribution is less than 1.3 , 1.0 and 1.9~\cm\
for the fundamental energy levels \n\1, \n\2 and \n\3 respectively.
With the growth in the size of the molecule
and increase in its total mass, the contribution of the
pseudo-potential to the vibrational energy levels decreases significantly
\cite{88GrPa.method}.  Therefore, to simplify and speed-up the
calculations we neglect the contribution of pseudo-potential. Some of
this contribution will be incorporate in our final, empirical
potential energy surface (PES).

Elements of the $\underline{G}(\underline{q})$  matrix are, in  general, a
complicated function of the vibrational coordinates \cite{88GrPa.method}. As an
example, Table~\ref{tab:gm1} gives elements of $\underline{G}(\underline{q})$
for the stretching and bending (inter-bond angle) modes. The latter in the case
of HNO\3\ are the angles $\angle$ N-O-N and N-O-H. The elements of
$\underline{G}(\underline{q})$ are presented for two cases: for the bending
coordinates given in the form of $\varphi = -\Delta \arccos \left( \frac{\vec
r_1 \vec r_2}{r_1 r_2} \right)$, and in the form of its cosine, $\varphi =
-\Delta \left( \frac{\vec r_1 \vec r_2}{r_1 r_2} \right)$. In total for HNO\3\
there are 55 elements of $\underline{G}(\underline{q})$.

\begin{table}[ht]
\scriptsize
\tabcolsep=3pt
\caption{Elements of the  $\underline{G}(\underline{q})$
matrix for two classes of nonlinear
bending coordinate $\varphi$, where $m_1$, $m_2$ and $m_3$
are the atomic masses, $r_1$ is the length of the bond between
atoms $1$ and $3$ and $r_2$ the length of the bond between atoms $2$ and $3$.}  \label{tab:gm1}
\begin{center}
\resizebox{\columnwidth}{!}{%
\begin{tabular}{c|c|c}
\hline\hline
&
$\varphi = -\Delta \arccos \left( \frac{\vec r_1 \vec r_2}{r_1 r_2} \right)$    &
$\varphi = -\Delta \left( \frac{\vec r_1 \vec r_2}{r_1 r_2} \right)$    \\
\hline
$g_{r_1 r_1}$    &
$\Bigl(\frac{1}{m_1} + \frac{1}{m_3}\Bigr)$    &
$\Bigl(\frac{1}{m_1} + \frac{1}{m_3}\Bigr)$    \\
$g_{r_2 r_2}$    &
$\Bigl(\frac{1}{m_2} + \frac{1}{m_3}\Bigr)$    &
$\Bigl(\frac{1}{m_2} + \frac{1}{m_3}\Bigr)$    \\
$g_{r_1 r_2} = g_{r_2 r_1}$    &
$\frac{\cos(\varphi)}{m_3}$    &
$\frac{\varphi}{m_3}$    \\

$g_{r_1 \varphi} = g_{\varphi r_1}$    &
$\frac{\sin(\varphi)}{r_2 m_3}$    &
$\frac{(1 - \varphi^2)}{r_2 m_3}$    \\

$g_{r_2 \varphi} = g_{\varphi r_2}$    &
$\frac{\sin(\varphi)}{r_1 m_3}$    &
$\frac{(1 - \varphi^2)}{r_1 m_3}$    \\

\multirow{2}{*}{$g_{\varphi \varphi}$}    &
$\frac{1}{r_1^2}\frac{m_1+m_3}{m_1 m_3} + \frac{1}{r_2^2}\frac{m_2+m_3}{m_2 m_3} +$    &
$\frac{(1-\varphi^2)}{r_1^2} \frac{m_1 + m_3}{m_1  m_3 } +
\frac{(1-\varphi^2)}{r_2^2}\frac{m_2+m_3}{m_2  m_3} +$    \\
&    $+\frac{2[(\cos(\varphi))^2 - \cos(\varphi)]}{(\sin(\varphi))^2 r_1 r_2 m_3}$
&    $+\frac{2(\varphi^2 - \varphi)}{r_1 r_2 m_3}$    \\
\hline
\end{tabular}
}
\end{center}
\end{table}

As can be seen from table~\ref{tab:gm1}, expressing the bending coordinates as a change in the angle
leads to a simplied expression for the matrix elements of
$\underline{G}(\underline{q})$ because their dependence on $\varphi$ is a
linear or quadratic. However, in general $\underline{G}(\underline{q})$ is a
complicated function of the internal coordinates. General expressions of its
elements are  given elsewhere \cite{88GrPa.method}. In general terms, if both
coordinates represent changes in bond lengths then
\begin{equation}
g_{ij}(\underline{q}) = g_{ij}^{0}(\underline{\varphi})~;
\end{equation}
if one coordinate represents a change in the bond length and the second is an angular
coordinate then
\begin{equation}
g_{ij}(\underline{q}) = \sum_{k} \frac{1}{r_k} g_{ij}^{k}(\underline{\varphi})~,
\end{equation}
and if both represent angular coordinates it becomes
\begin{equation}
g_{ij}(\underline{q}) = \sum_{k,l} \frac{1}{r_k r_l} g_{ij}^{kl}(\underline{\varphi}) ~.
\end{equation}
In these expressions $r_k$ is bond length of the $k$-th bond and $\underline{\varphi}$
represents the angular coordinates.

$\underline{G}(\underline{q})$
is computed using a second-order Taylor expansion in the angular coordinates
\begin{equation}
\label{e:g0}
g_{ij}(\underline{q}) = g_{ij}^{0}(0)
     + \sum_{m} \left( \frac{\partial g_{ij}^{0}(\underline{\varphi})}{\partial \varphi_m} \right)_0 \varphi_m
     + \frac{1}{2} \sum_{m,n}
       \left( \frac{\partial^2 g_{ij}^{0}(\underline{\varphi})}{\partial \varphi_m \partial \varphi_n} \right)_0 \varphi_m \varphi_n ~~~,
\end{equation}
for the case where  $i$ and $j$ both represent bonds;
\begin{equation}
\label{e:g1}
g_{ij}(\underline{q}) = \sum_{k} \frac{1}{r_k}
       \left[ g_{ij}^{k}(0)
     + \sum_{m} \left( \frac{\partial g_{ij}^{k}(\underline{\varphi})}{\partial \varphi_m} \right)_0 \varphi_m
     + \frac{1}{2} \sum_{m,n}
       \left( \frac{\partial^2 g_{ij}^{k}(\underline{\varphi})}{\partial \varphi_m \partial \varphi_n} \right)_0 \varphi_m \varphi_n \right] ~~~,
\end{equation}
for the case of one bond and one bond length;
\begin{equation}
\label{e:g2}
g_{ij}(\underline{q}) = \sum_{k,l} \frac{1}{r_k r_l}
       \left[ g_{ij}^{kl}(0)
     + \sum_{m} \left( \frac{\partial g_{ij}^{kl}(\underline{\varphi})}{\partial \varphi_m} \right)_0 \varphi_m
     + \frac{1}{2} \sum_{m,n}
       \left( \frac{\partial^2 g_{ij}^{kl}(\underline{\varphi})}{\partial \varphi_m \partial \varphi_n} \right)_0 \varphi_m \varphi_n \right] ~.
\end{equation}
for two angles.
It is beneficial to choose an internal coordinate
in $\underline{\varphi}$  as cosine differences $q_i = \cos
\varphi_i - \cos \varphi_i^e$, where $\cos \varphi_i^e$ is the instantaneous
equilibrium angle for bond angles. As can be seen from Table~\ref{tab:gm1},
these expansions in terms of $\underline{\varphi}$  are exact as
$g_{ij}^{0}(\underline{\varphi})$, $g_{ij}^{k}(\underline{\varphi})$ and
$g_{ij}^{kl}(\underline{\varphi})$ are quadratic functions of the angular
coordinates. In this expansion, a sine difference  $q_i = \sin \varphi_i - \sin
\varphi_i^e$ is the obvious choice for an internal coordinate describing a dihedral mode.

To simplify the calculation of the vibrational Hamiltonian matrix elements in
block 2 (see below), which correspond to the perturbative contribution for the
matrix elements from the main block 1 (see below), the vibrational kinetic
energy coefficients are expanded in the polynomial form and truncated at the
second order
\begin{equation}
g_{ij}(\underline{q}) = g_{ij}(0)
     + \sum_{m} \left( \frac{\partial g_{ij}(\underline{q})}{\partial q_m} \right)_0 q_m
     + \frac{1}{2} \sum_{m,n}
       \left( \frac{\partial^2 g_{ij}(\underline{q})}{\partial q_m \partial q_n} \right)_0 q_m q_n ~.
\end{equation}
This form is convenient because it allows faster, by an order-of-magnitude or more,
computation of the coefficients without significant loss of accuracy.

The potential energy function used by us is a fourth-order polynomial
\begin{equation}
\label{e:potential}
V(\underline{q}) = \frac{1}{2}  \sum_{i,j} D_{ij} x_i x_j
     + \frac{1}{6}  \sum_{i,j,k} D_{ijk} x_i x_j x_k
     + \frac{1}{24} \sum_{i,j,k,l} D_{ijkl} x_i x_j x_k x_l ~,
\end{equation}
where Morse coordinates, $x_i = (1-\exp^{-\alpha_i \Delta r_i})$, are used to
represent  changes of all bonds for HNO\3 molecule and $x_i = q_i$ for angular
coordinates.

\section{Method}
\label{s:Hybrid}

For ease of use and better convergence of
the basis functions, the $\chi_i$ are generally chosen to form a complete
orthonormal set.
In variational calculations of  vibrational energy levels, the Hamiltonian matrix elements
%Eq.~(\ref{e:hmtr})
are computed using the product form
\begin{equation}
\label{e:basis}
\chi_{kn} = \prod_{i} \phi_{k_{i}}(r_i) \prod_{s} \psi_{n_{s}}(Q_s)
\end{equation}
of the basis functions, which are eigenfunctions of the Morse or harmonic oscillators.
Morse oscillator functions, $\phi_{k_{i}} (r_i)$, are used for
 the stretching coordinates, $r_i$, for which the potential is given using a Morse
coordinate. Harmonic basis functions, $\psi_{n_{s}} (Q_s)$, are used for
the other coordinates, which are represented using curvilinear normal
coordinates
\begin{equation}
\label{e:q-normal}
Q_s = \sum_{i} L^{q}_{is} q_i
\end{equation}
expressed as a linear sum over the internal coordinates, $q_i$, for which the potential
function is  defined as a Taylor series. The coordinates $Q_s$ are those which
diagonalize the harmonic part of the Hamiltonian given in the internal coordinates
$q_i$. With these definitions, all multi-dimensional integrals required to calculate the
Hamiltonian matrix elements  are separated into products of one-dimensional integrals
between either Morse functions or harmonic oscillators. All these integrals have a
simple analytic form which results in high-speed computation of the Hamiltonian matrix
elements.

The vibrational Hamiltonian matrix constructed in this way is then diagonalized
to give the vibrational energy levels $E^{vib}_\lambda$ and the corresponding
wave functions $\phi^{vib}_\lambda$.

Our implementation relies on the particular structure of the
Hamiltonian matrix ordered by increasing polyad (total vibrational
excitation) number, $N_V$
\begin{equation}
\label{e:N_V}
 N_V =   \sum_ {m = 1}^{N_c} a_m v_m
\end{equation}
where $a_m$ is some weighting which is often roughly proportional to
the inverse of the frequency \cite{trove-paper}. For simplicity
in this work we use $a_m = 1$ for all $m$.
This gives the size of the basis set, $M^{\rm max}_B$ in terms of the maximum
polyad number, $N_V^{\rm max}$,
\begin{equation}
\label{e:N_Bmax}
M^{\rm max}_B = \frac{(N_V^{\rm max} + N_c)!}{N_V^{\rm max}! N_c!} =
\prod_{i=1}^{N_c} (N_V^{\rm max} + i)/i .
\end{equation}

Calculating all $N_V \le N_{V}^{\rm target}$ vibrational term values
for HNO\3 with an accuracy better than 0.3 cm$^{-1}$ requires basis
functions with $N_V^{\rm max} \geq N_{V}^{\rm target} +9$. This means that
the Hamiltonian matrix must include all the basis functions for which the
difference in $v_m$ is larger than 9.
Thus,  an accurate calculation of the fourth overtones and combination
frequencies ($N_{V}^{\rm target} = 5$) demands a variational basis for HNO\3 ($N_c = 9$)
which includes $M_B^{\rm max} = 817~190$ basis functions.

\begin{figure}[t!]
\centering
\includegraphics[width = 0.4\textwidth]{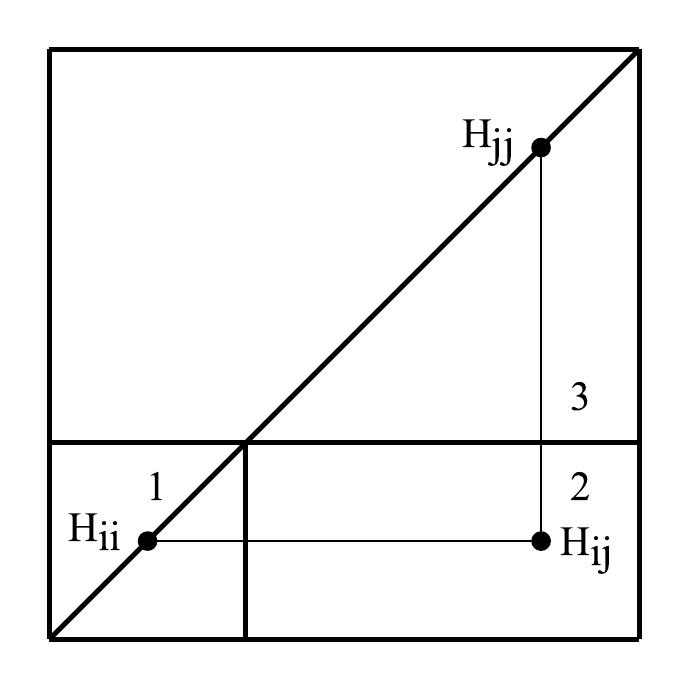}
\caption{Block structure of the vibrational Hamiltonian matrix.
    Region 1: matrix elements with the largest contributions to the
    target energy levels ($N_V \leq N_{v}^{\rm target} + 4$);
    Region 2 contains elements with small contributions to these states.
    The contribution from elements in Region 3 is disregarded.}
\label{fig:1}
\end{figure}

We use a hybrid variational-perturbation method for calculating ro-vibrational energy
levels of a polyatomic molecules \cite{jt588}.
It combines the advantages of both variational calculations and perturbation theory.
The vibrational problem is solved by diagonalizing a Hamiltonian matrix, which is
partitioned into two sub-blocks, as shown in Fig.~\ref{fig:1}.
The first, smaller sub-block includes matrix elements with the largest contribution
to the energy levels targeted in the calculations ($N_V \leq N_{v}^{\rm target} + 4$).
The second, larger sub-block comprises those basis states which have little effect on these
energy levels. Numerical perturbation theory, implemented as a Jacobi rotation,
is used to compute the contributions from the matrix elements of the second sub-block.
Only the first sub-block needs to be stored in memory and diagonalized.
The size of block 1, $M_B^{(1)}$,
is given by Eq.~(\ref{e:N_Bmax}) using
$N_V^{(1)}$ as the number of the largest polyad included in block 1.

In the first step of our approach all the matrix elements from blocks
1 are computed along with the diagonal matrix elements of block 3.
The second step involves computing the off-diagonal
elements of block 2 and accounting for their effect on the matrix
elements of the block 1 using one Jacobi rotation
\cite{1846Ja.method,1965Wi.method}. Considering a contribution from
off-diagonal element $H_{ij}$ in block 2, which couples the diagonal
elements $H_{ii}$ in block 1 and $H_{jj}$ in block 3, the
diagonal element in block 1 is perturbatively adjusted using the Jacobi formula
\begin{equation}
\tilde{H}_{ii} = H_{ii} - \sum_{j \in {\rm block 2}} \Delta E_{ij} ~ ,
\end{equation}
where
\begin{equation}
\Delta E_{ij} = \frac{{\rm sign}(\sigma_{ij}) H_{ij} }{|\sigma_{ij}| + \sqrt{1 + \sigma_{ij}^2}}
\end{equation}
and
\begin{equation}
\sigma_{ij} = \frac{(H_{jj} - H_{ii})}{ 2 H_{ij} } ~.
\end{equation}

For ro-vibrational energy levels it is necessary to calculate elements of the
complex Hermitian Hamiltonian matrix
\begin{equation}
H^{JJ\p}_{\lambda k m, \lambda\p k\p m\p} =
\langle \chi^{J}_{\lambda k m} \arrowvert \hat H_{vr} \arrowvert
\chi^{J\p}_{\lambda\p k\p m\p} \rangle ~,
\end{equation}
using the basis functions
\begin{equation}
\chi^{J}_{\lambda km} = \Phi^{vib}_{\lambda} \phi^{J}_{km} ~~~,~~~
\phi^{J}_{km} = \left( \frac{2J+1}{8 \pi^2} \right)^{1/2} D^{J~*}_{km},
\end{equation}
where $D^{J}_{km}$ is a  (complex) Wigner function. In this case
\begin{equation}
\label{e:H-matrix}
H^{J}_{\lambda k, \lambda\p k\p } =
E^{vib}_\lambda \delta_{\lambda \lambda\p} \delta_{kk\p}
- \frac{\hbar^2}{2} \sum_{a,b} \bar{\mu}^{\lambda \lambda\p}_{ab}
  \langle \phi^{J}_{km} \arrowvert \frac{\partial^2}{\partial \xi_a \partial \xi_b}
  \arrowvert \phi^{J}_{k\p m} \rangle.
\end{equation}
The vibrationally averaged moment of interia, $\bar{\mu}^{\lambda \lambda\p}_{ab}$, is expanded to second-order as a Taylor series
\begin{equation*}
\bar{\mu}^{\lambda \lambda\p}_{ab} =
\langle \Phi^{vib}_{\lambda} \arrowvert \mu_{ab}(\underline{q})
\arrowvert \Phi^{vib}_{\lambda\p} \rangle =
\end{equation*}
\begin{equation}
\label{e:M-matrix}
\mu_{ab}(0) \delta_{\lambda \lambda\p} +
\sum_{m} \left( \frac{\partial \mu_{ab}(\underline{q})}{\partial q_m} \right)_0
    \langle \Phi^{vib}_{\lambda} \arrowvert  q_m \arrowvert \Phi^{vib}_{\lambda\p} \rangle +
\frac{1}{2} \sum_{m,n}
  \left( \frac{\partial^2 \mu_{ab}(\underline{q})}{\partial q_m \partial q_n} \right)_0
    \langle \Phi^{vib}_{\lambda} \arrowvert q_m q_n \arrowvert \Phi^{vib}_{\lambda\p} \rangle
\end{equation}
and all the integrals reduce to products of one-dimensional integrals over
either Morse or harmonic oscillators.

The off-diagonal elements of the vibration-rotation Hamiltonian matrix
\begin{equation}
H^{J}_{\lambda k, \lambda\p k\p} =
- \frac{\hbar^2}{2} \sum_{a,b} \bar{\mu}^{\lambda \lambda\p}_{ab}
  \langle \phi^{J}_{km} \arrowvert \frac{\partial^2}{\partial \xi_a \partial \xi_b}
  \arrowvert \phi^{J}_{k\p m\p} \rangle \delta_{JJ\p}~,
\end{equation}
differ significantly in magnitude, depending on whether they are diagonal
in the vibrations, $\lambda = \lambda\p$, or couple different vibrational states,
$\lambda \ne \lambda\p$. 

When calculating the vibrational-rotational energy levels, the
off-diagonal elements $H^{J}_{\lambda k, \lambda\p k\p}$ corresponding
to different vibrational states $\lambda \ne \lambda\p$ give a much
smaller contribution (change in the diagonal elements in the block
that will be diagonalized) to the calculated energy levels than the
off-diagonal elements $H^{J}_{\lambda k, \lambda k\p}$ within the
vibrational state in question.  These changes are given approximately by
\begin{equation}
\Delta E_{\lambda \lambda\p} = \frac{\left({H^{J}_{\lambda k, \lambda\p k\p}}\right)^{2}}
{H^{J}_{\lambda\p k\p, \lambda\p k\p} - H^{J}_{\lambda k, \lambda k}},
\end{equation}
\begin{equation}
\Delta E_{\lambda \lambda} = \frac{\left({H^{J}_{\lambda k, \lambda k\p}}\right)^{2}}
{H^{J}_{\lambda k\p, \lambda k\p} - H^{J}_{\lambda k, \lambda k}} ~.
\end{equation}
However,
\begin{equation}
|H^{J}_{\lambda\p k\p, \lambda\p k\p} - H^{J}_{\lambda k, \lambda k}| \gg
|H^{J}_{\lambda k\p, \lambda k\p} - H^{J}_{\lambda k, \lambda k}|
\end{equation}
since $(H^{J}_{\lambda k\p, \lambda k\p} - H^{J}_{\lambda k, \lambda k})$
involves only a change in the rotational energy level, while $(H^{J}_{\lambda\p
k\p, \lambda\p k\p} - H^{J}_{\lambda k, \lambda k})$ involves also a change in
the vibrational energy level. For semi-rigid molecules with small values of
the vibrational quantum numbers the following condition usually holds
\begin{equation}
|H^{J}_{\lambda k, \lambda k\p }| \approx |H^{J}_{\lambda k, \lambda\p k\p }| ~~~,~~~ \lambda \ne \lambda\p,
\end{equation}
which results from the slight change in the effective geometry of the molecule
upon vibrational excitation. This feature of the vibrational-rotational
Hamiltonian matrix is common for large molecules and underpins the
ro-vibrational version of our hybrid approach \citep{jt588}. Again we use
second-order perturbation theory, as defined by a Jacobi rotation, to transform
the $H^{J}_{\lambda k,\lambda\p k\p}$ matrix to a series of much smaller
rotational sub-matrices corresponding to different vibrational states
$\lambda$, $\tilde{H}^{J}_{\lambda k,\lambda k\p}$. The dimension of each
rotational sub-block is $(2J+1)$ only and we consider $M_B^{\rm target}$
sub-matrices that correspond to  $M_B^{\rm target}$ vibrational states.

As above, we employ a single Jacobi rotation which we apply to the
ro-vibrational Hamiltonian.  The best agreement with the variational solution
is achieved when both the diagonal and off-diagonal elements are updated
\cite{jt588} as given by
\begin{equation}
\label{e:H-diag}
\tilde{H}^{J}_{\lambda k,\lambda k} = H^{J}_{\lambda k,\lambda k} + \sum_{ \lambda\p \in M_B^{\rm vib}, \lambda\p \ne \lambda} \sum_{k\p} t_{\lambda k,\lambda\p k\p} \eta_{\lambda k,\lambda\p k\p}
\end{equation}
for the diagonal elements
\begin{equation}
\label{e:H-nondiag}
\begin{split}
\tilde{H}^{J}_{\lambda k,\lambda k\pp} =
\frac{1}{2} \sum_{\lambda\p \in M_B^{\rm vib}, \lambda\p \ne \lambda} \sum_{k\p}
& \left[c_{\lambda k,\lambda\p k\p} H^{J}_{\lambda k,\lambda k\pp} + c_{\lambda k\pp,\lambda\p k\p} H^{*J}_{\lambda k,\lambda k\pp} \right. + \\
& \left. + s_{\lambda k,\lambda\p k\p} H^{J}_{\lambda\p k\p,\lambda k\pp} + s_{\lambda\p k\p,\lambda k\pp H^{J}_{\lambda k,\lambda\p k\p}} \right],
\end{split}
\end{equation}
which is a symmetrized version of the standard
formula for the single Jacobi rotation with respect to the indices $\lambda k$
and $\lambda k\pp$.
For the off-diagonal elements
\begin{eqnarray}
\nonumber
c_{\lambda k,\lambda\p k\p} &=& \frac{1}{\sqrt{1 + t_{\lambda k,\lambda\p k\p}^2}} ~, \\
\nonumber
s_{\lambda k,\lambda\p k\p} &=& \frac{c_{\lambda k,\lambda\p k\p} t_{\lambda k,\lambda\p k\p}}{\eta_{\lambda k,\lambda\p k\p}} H^{J}_{\lambda k,\lambda\p k\p} ~, \\
\nonumber
t_{\lambda k,\lambda\p k\p} &=& {\rm sign} (\vartheta_{\lambda k,\lambda\p k\p}) / (|\vartheta_{\lambda k,\lambda\p k\p}| + \sqrt{1 + \vartheta_{\lambda k,\lambda\p k\p}^2}) ~~~,~~~ \\
\nonumber
\eta_{\lambda k,\lambda\p k\p} &=& {\rm sign}\left[{\operatorname{Re} (H^{J}_{\lambda k,\lambda\p k\p} ) } \right] |H^{J}_{\lambda k,\lambda\p k\p}| ~~~, \\
\nonumber
\vartheta_{\lambda k,\lambda\p k\p} &=& \frac{H^{J}_{\lambda k,\lambda k} - H^{J}_{\lambda\p k\p,\lambda\p k\p}}{2 \eta_{\lambda k,\lambda\p k\p}} ~~~,
\end{eqnarray}
where $\underline{H}^{J}$ and $\underline{\tilde{H}}^{J}$ are the initial
(unperturbed) matrix and perturbed matrix, respectively; $\lambda$ runs from 1
to $M_B^{\rm target}$, and $k = -J \ldots +J$.

The resulting block-diagonal form is then diagonalized for each $\lambda$
sub-matrix separately. Thus our algorithm
replaces the diagonalization of a huge [$M_B^{\rm vib}\times (2J+1)$]-dimensional
ro-vibrational matrix with a number of diagonalizations of much smaller
dimenionsional-[$(2J+1)$] matrices.

\section{Potential energy and dipole moment functions}
\label{s:Potential}

An important factor in solving the anharmonic vibrational problem is the choice
of internal curvilinear vibrational coordinates, $q_i$.  This choice determines
how close the truncated polynomial potential function of
Eq.~(\ref{e:potential}) is to the real PES of the molecule, as well as the
specific form of the kinetic energy coefficients matrix,
$\underline{G}(\underline{q})$, see Eqs.~(\ref{e:g0}), (\ref{e:g1}) and
(\ref{e:g2}).

\begin{figure}[t!]
\centering
\includegraphics[width=0.4\textwidth]{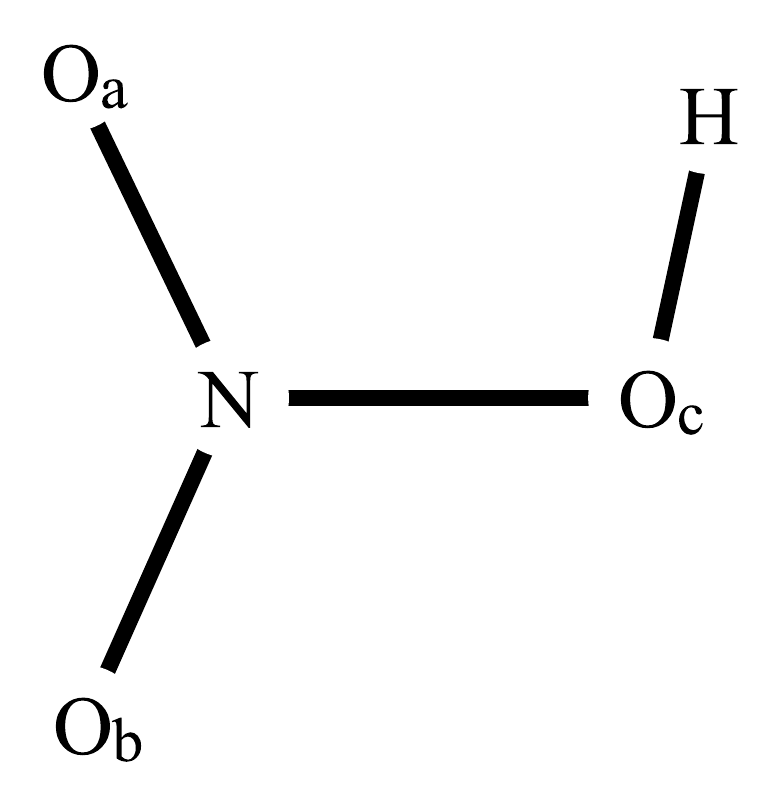}
\caption{Structure of the HNO\3 molecule.}
\label{fig:3}
\end{figure}

The structure of the  HNO\3 molecule and the atom numbering we use is shown in Fig.~\ref{fig:3}.
For HNO\3 we employ the following vibrational coordinates:
\begin{itemize}
\item Four coordinates represent changes in the length of valence bonds
    between atoms
$$
\Delta r_i = \left\{ \Delta r_{{\rm NO}_a} , \Delta r_{{\rm NO}_b} , \Delta r_{{\rm NO}_c} , \Delta r_{{\rm O}_c{\rm H}} \right\} ~,
$$
where
$$
r_i = \left\{ r_{{\rm NO}_a} , r_{{\rm NO}_b} , r_{{\rm NO}_c} , r_{{\rm O}_c{\rm H}} \right\}
$$
are lengths of the bond vectors
$$
\vec r_i = \left\{ \vec r_{{\rm NO}_a} , \vec r_{{\rm NO}_b} , \vec r_{{\rm NO}_c} , \vec r_{{\rm O}_c{\rm H}} \right\}
$$
with the vectors pointing from the first to the second atom.
\item
Four coordinates represent changes  in the cosines of angles between the valence bonds
$$
\varphi_m = \left\{
-\Delta \left( \frac{\vec r_{{\rm NO}_a} \vec r_{{\rm NO}_b}}{r_{{\rm NO}_a} r_{{\rm NO}_b}} \right) ,
-\Delta \left( \frac{\vec r_{{\rm NO}_a} \vec r_{{\rm NO}_c}}{r_{{\rm NO}_a} r_{{\rm NO}_c}} \right) ,
-\Delta \left( \frac{\vec r_{{\rm NO}_b} \vec r_{{\rm NO}_c}}{r_{{\rm NO}_b} r_{{\rm NO}_c}} \right) ,
-\Delta \left( \frac{\vec r_{{\rm NO}_c} \vec r_{{\rm O}_c{\rm H}}}{r_{{\rm NO}_c} r_{{\rm O}_c{\rm H}}} \right) \right\} ~ .
$$
which contain one dependent or redundant angle. This dependence is
removed upon the introduction
of $Q_s$ coordinates, see Eq.~(\ref{e:q-normal}).
\item A coordinate corresponding to the change in the sine of the angle
    between the orientation of bond ${\rm NO}_c$ from the plane formed by
    bonds ${\rm NO}_a$ and ${\rm NO}_b$
$$
\varphi_f = \Delta \left[ \frac{\vec r_{{\rm NO}_c} (\vec r_{{\rm NO}_a} \times \vec r_{{\rm NO}_b})}
{r_{{\rm NO}_c} r_{{\rm NO}_a} r_{{\rm NO}_b}} \right] ,
$$
\item A coordinate corresponding to the change in the sine of the angle
    between the plane formed by bonds ${\rm NO}_a$, ${\rm NO}_b$ and the
    plane formed by bonds ${\rm NO}_c$, ${\rm O}_c{\rm H}$ while rotating
    them relative to each other around ${\rm NO}_c$ bond
$$
\varphi_\rho = \Delta \left\{ \frac{\vec \rho \left[ (\vec r_{{\rm NO}_a} \times \vec r_{{\rm NO}_b}) \times (\vec r_{{\rm NO}_c} \times \vec r_{{\rm O}_c{\rm H}}) \right]}
{r_{{\rm NO}_a} r_{{\rm NO}_b} r_{{\rm NO}_c} r_{{\rm O}_c{\rm H}}} \right\} .
$$
where $\vec \rho$ is a unit vector perpendicular to the equilibrium plane of the molecule.
\end{itemize}

This physically-motivated choice of internal curvilinear vibrational
coordinates ensures that quartic expansion of the potential,
Eq.~(\ref{e:potential}), stays close to the real PES in the region around the
minimum.  For example, HNO\3 has a low frequency mode, \n\9, corresponding to
rotation of the ${\rm O}_c{\rm H}$ bond around the ${\rm NO}_c$ bond which is
approximately represented by the coordinate $\varphi_\rho$. This vibrational
mode is characterized by very large anharmonicity, as it has large amplitude
and a strongly anharmonic potential function. Quantum-chemical calculations
carried out by Lauvergnat and Nauts \cite{10LaNaxx.HNO3} show that this
rotation corresponds to the potential curve which is close to a sine wave.  Our
quantum-chemical calculations show that in the expansion of the PES as a
function of coordinate $\varphi_\rho$ in Eq.~(\ref{e:potential}) gives a
nonzero quadratic term for coordinate $\varphi_\rho$ but zero cubic and quartic
terms.  This is because writing the potential function in the form
$V(\underline{\varphi_\rho}) = \frac{1}{2} D_{\varphi_\rho
\varphi_\rho} \varphi_\rho^2$ is equivalent to defining
it as a sine wave in the angle of rotation of the ${\rm O}_c{\rm H}$ bond.
Thus, choosing the vibrational coordinates in the form of change of sine of the
angle for this mode provides a compact definition of the strongly anharmonic
sinusoidal potential function as a single term in the expansion
Eq.~(\ref{e:potential}).  As a result, we obtain good agreement between the
experimental and calculated vibrational terms values for $n$\n\9 using our
{\it ab initio} potential  parameter, see Table~\ref{tab:n9}. This in turn 
means that the calculated spectrum reproduces the absorption intensities for
the hot bands starting from the \n\9 , 2\n\9 , 3\n\9 states which are shifted
from the main absorption bands and which are characteristic of HNO\3,
see Figure~\ref{fig:h2} and discussion below. 

\begin{table}[ht]
\scriptsize
\tabcolsep=3pt
\caption{Experimental (Exp.) \cite{94PeFlCA.HNO3} and calculated (Calc.)
vibrational term values for excitation of the \n\9 mode
using {\it ab initio} potential parameters; the result of
Lauvergnat and Nauts \cite{10LaNaxx.HNO3} are given for comparison} \label{tab:n9}
\begin{center}
\begin{tabular}{rrrr}
\hline\hline
Transition & \multicolumn{3}{c}{Frequency (\cm)} \\
\cline{2-4}
      &   Exp.   & Calc. \cite{10LaNaxx.HNO3} & Our calc. \\
\hline
 \n\9 &   458.2  &   458.0  &   456.7  \\
2\n\9 &   896.3  &   886.6  &   894.1  \\
3\n\9 &  1289.0  &  1293.1  &  1284.2  \\
4\n\9 &  1664.7  &  1671.9  &  1656.7  \\
\hline\hline
\end{tabular}
\end{center}
\end{table}

We use a simplified method for the initial calculation of parameters for the PES
and dipole moment function (DMF). This simplification is justified because our
neglect of the pseudo-potential as well as other small contributions
such as adiabatic effects \cite{jt309} means that we cannot calculate
energy levels {\it ab initio} with the accuracy needed to compute a
final line list. Therefore, it is necessary to improve the PES and DMF
by solving the inverse spectral problem.

First, we  compute a force constant expansion for the PES
using a fourth-order Taylor series
\begin{equation}
\label{e:begpot}
V(\underline{q}) = \frac{1}{2}  \sum_{i,j} f^{0}_{ij} q_i q_j
     + \frac{1}{6}  \sum_{i,j,k} f^{0}_{ijk} q_i q_j q_k
     + \frac{1}{24} \sum_{i,j,k,l} f^{0}_{ijkl} q_i q_j q_k q_l
\end{equation}
in internal  curvilinear  coordinates $ q_i$. Initial values for the quadratic
force constants $f^{0}_{ij}$ were calculated using the central finite
difference relations
\begin{equation}
f_{ii}^0 = \frac {E^{+} + E^{-} - 2 E^{0}} {\delta q_i^2} ~~~ , ~~~
f_{ij}^0 = \frac {E^{++} + E^{--} - E^{+-} - E^{-+}}{4 \delta q_i \delta q_j}
\end{equation}
where $E^{0}$ is the molecular energy of the equilibrium configuration  and
$E^{\pm}$ is the energy of the geometry in which  the vibrational coordinate
$q_i$ is increased/decreased by ${\delta q_i}$. $E^{++}$, $E^{--}$, $E^{+-}$,
$E^{-+}$ are energies for geometries in which the vibrational coordinates $q_i$
and $q_j$ are increased and/or decreased by multiples of ${\delta q_i}$ and
${\delta q_j}$. Energies and the equilibrium geometry of HNO\3 were calculated
{\it ab initio\/} at the CCSD(T)/aug-cc-pVQZ level of theory using MOLPRO
\cite{12WeKnKn.methods}. Similarly, the initial values of cubic $f^{0}_{ijk}$
and quartic $f^{0}_{ijkl}$ force constants were calculated using the finite
difference relations
\begin{equation}
f_{ijk}^0 = \frac {f_{ij}^{+} + f_{ij}^{-} - 2 f_{ij}^{0}} {\delta q_k^2} ~~~ , ~~~
f_{ijkl}^0 = \frac {f_{ij}^{++} + f_{ij}^{--} - f_{ij}^{+-} - f_{ij}^{-+}}{4 \delta q_k \delta q_l}
\end{equation}
where $f_{ij}^{0}$ are the Hessian (quadratic force constants) at the
equilibrium geometry, $f_{ij}^{\pm}$ is the Hessian corresponding to an
increase/decrease in the vibrational coordinates $q_i$ by ${\delta q_i}$. The
Hessians $f_{ij}^{++}$, $f_{ij}^{--}$, $f_{ij}^{+-}$, $f_{ij}^{-+}$ are
computed at geometries obtained by increasing and decreasing vibrational
coordinates $q_i$ and $q_j$ by multiples of ${\delta q_i}$ and ${\delta q_j}$.
These Hessians were calculated {\it ab initio\/}  at the MP2/aug-cc-pVQZ level
of theory using Gaussian \cite{g09.method}. These calculations were based on
the MP2/aug-cc-pVQZ equilibrium geometry.

The second step of the calculation uses the initial force constants
$f^{0}_{ij}$, $f^{0}_{ijk}$, $f^{0}_{ijkl}$  to construct the constants
$D_{ij}$ , $D_{ijk}$ , $D_{ijkl}$ which are used to represent the PES,
see Eq.~(\ref{e:begpot}).
This requires taking into account the relation between the $f$, $D$ and $\alpha$
constants for the Morse oscillator, for example, $f^{0}_{ii} = D_{ii} \alpha_i^2$.

In the third stage of the calculation the PES parameters are refined using the empirical
values of the energy levels. Solution of the inverse spectral problem is facilitated by
the analytic evaluation of the first derivatives of the energy levels, $E_i$, with
respect to PES parameters $D_j$ using the Hellmann-Feynman theorem:
\begin{equation}
\frac{\partial E_i}{\partial D_j} =
\langle \psi_i \arrowvert \frac{\partial \hat H_v}{\partial D_j} \arrowvert \psi_i \rangle =
\langle \psi_i \arrowvert \frac{\partial V(\underline{q})}{\partial D_j} \arrowvert \psi_i \rangle
\end{equation}
where $\psi_i$ is the wave function of  energy level $E_i$.

Our solution of the inverse spectral problem is based on the method of
regularization due to Tikhonov \cite{79TiArxx.method, 92GrDexx.method}. This
method minimizes the functional
\begin{equation}
\Phi = \sum_{i}\left (E^c_i - E^e_i \right)^2 W_i +
\sum_{j}\left[ \alpha_j \left( \frac{D_j-D^{0}_j}{\Delta D_{j}^{max}} \right)^2 +
               \beta_j  \left( \frac{D_j-D^{0}_j}{\Delta D_{j}^{max}} \right)^{10} \right].
\end{equation}
where $E^c_i$ , $E^e_i$ and $W_{i}$ are the calculated and experimental values
of the energy level and its weight, $D_j$ and $D^{0}_j$ are the current and
initial values of the parameters, and $\Delta D_{j}^{max}$ is the maximum
possible deviation of parameter value from its initial value. In this formula,
$\alpha_j$ and $\beta_j$ are regularization parameters that allow one to control
progress in solving the inverse problem.
The terms containing $D_j$ in this functional allows one to
constrain the refined parameters to their initial ({\it ab initio}) values.
The method of regularization ensures that there is always a valid solution, even
when the number of experimental energies is less than the number of variable
parameters. This is similar to the method where the shape of the potential
functions are controlled by constraining directly to the {\it ab initio}
energies (see, for example, \citep{03YuCaJe.PH3}).

The dipole moment of HNO\3 is represented as a second-order polynomial
\begin{equation}
\label{e:dipol}
\vec D(\underline{q}) = \vec D^{0} + \sum_{i} \vec d_{i} q_i
+ \frac{1}{2}  \sum_{i,j} \vec d_{ij} q_i q_j,
\end{equation}
where $\vec D^{0}$ is the equilibrium value of the dipole moment, $\vec d_{i}$
and $\vec d_{ij}$ equal, respectively, the first and second derivatives of the
dipole moment with respect to the curvilinear coordinates $q_i$ and $q_j$.
Initial values of $\vec d_{i}$ and $\vec d_{ij}$ were calculated using the
finite difference relations
\begin{equation}
\vec d_{i} = \frac {\vec D^{+} + \vec D^{-} - 2 \vec D^{0}} {\delta q_i^2} ~~~ , ~~~
\vec d_{ij} = \frac {\vec D^{++} + \vec D^{--} - \vec D^{+-} - \vec D^{-+}}{4 \delta q_i \delta q_j},
\end{equation}
where $\vec D^{\pm}$ is the dipole moment corresponding to an increase/decrease
in the vibrational coordinates $q_i$ by ${\delta q_i}$. Dipole moments $\vec
D^{++}$, $\vec D^{--}$, $\vec D^{+-}$, $\vec D^{-+}$ correspond to the
geometries obtained by  increasing or decreasing vibrational coordinates $q_i$
and $q_j$ by multiples of ${\delta q_i}$ and ${\delta q_j}$. Dipole moments
were calculated {\it ab initio}  at the CCSD(T)/aug-cc-pVQZ level of theory
using MOLPRO from the change in energy of the molecule in an external electric
field, which is considered the better of the methods for computing {\it ab
initio} dipoles \cite{jt475}.

Calculation of initial values of the PES and DMF parameters and all subsequent
calculations were made using program ANGMOL \cite{88GrPa.method}.  ANGMOL
automatically produces the necessary inputs for MOLPRO and Gaussian, runs these
programs and extracts the required data (energy, Hessian and dipole moment)
from their listings. These calculations take into account the  relationship
between our internal curvilinear coordinates, $q_i$, and the Cartesian
coordinates of the atoms \cite{88GrPa.method}.

The calculated initial values of the force constants and the dipole moment of the molecule
depends on the increment, ${\delta q_i}$, used to evaluate the derivatives.
For small values of ${\delta q_i}$, these derivatives are distorted by the finite numerical
precision inherent in Gaussian and MOLPRO, while for large increments the PES and DMF
may not be quadratic. We find that optimal increments, ${\delta q_i}$, are
0.01~\AA\ for bonds and 0.01 for changes in  cosines and sines of the angular coordinates.
These increments were used in all calculations of the initial PES and DMF coefficients.

Ro-vibrational calculations showed that the CCSD(T)/aug-cc-pVQZ
equilibrium geometry does not accurately describe the rotational
energy levels. Therefore we used a modified geometry in which all bond
lengths were reduced by 0.1~\%.  Table~\ref{tab:geom} gives our
calculated and modified equilibrium geometry.

\begin{table}[ht]
\scriptsize
\tabcolsep=3pt
\caption{Equilibrium bond lengths (\AA) and angles (\degree) in the molecule HNO\3 which is planar.} \label{tab:geom}
\begin{center}
\begin{tabular}{lrr}
\hline\hline
Parameter              & {\it Ab initio}   & Modified  \\
\hline
$r_{{\rm NO}_a}$             &   1.21032    &   1.20911    \\
$r_{{\rm NO}_b}$             &   1.19531    &   1.19412    \\
$r_{{\rm NO}_c}$             &   1.39960    &   1.39820    \\
$r_{{\rm O}_c{\rm H}}$             &   0.96985    &   0.96888    \\
$\alpha_{{\rm NO}_a,{\rm NO}_b}$   &   130.2713   &   130.2713   \\
$\alpha_{{\rm NO}_a,{\rm NO}_c}$   &   115.7199   &   115.7199   \\
$\alpha_{{\rm NO}_b,{\rm NO}_c}$   &   114.0088   &   114.0088   \\
$\alpha_{{\rm NO}_c,OH}$     &   102.2040   &   102.2040   \\
\hline\hline
\end{tabular}
\end{center}
\end{table}

Finally we note that the program ANGMOL is freely available on upon request
to the first author.

\section{Calculated vibrational term values}
\label{s:Vibrational}

First the vibrational energy levels were calculated using the Hamiltonian described
in Section~\ref{s:hamiltonian} and our hybrid variational-perturbation method \cite{jt588},
as implemented in ANGMOL \cite{88GrPa.method}.

An important feature of the HNO\3 IR spectrum \cite{94PeFlCA.HNO3,
04FeHaVa.HNO3, pnnl} in the 0 - 7000~\cm\ range is that absorption is dominated
by the fundamental (\n$_i$), first overtones (2\n$_i$), and first combination
(\n$_i$ +\n$_j$) bands. The presence of low-frequency vibrations with high
anharmonicity also leads to observation of hot band transitions, such as
4\n$_i$-3\n$_i$ and 4\n$_i$-2\n$_i$, even at room temperature. The strongest
hot-band transitions correspond to those involving the low-frequency \n\9 mode.
In addition, the spectrum is further complicated by strong Fermi resonances,
for example between \n\5 and 2\n\9.

We aim to make accurate calculations (better than 1~\cm) for vibrational
states with the quantum numbers up to
$N_{V}^{\rm target} = 5$ which corresponds to
$M_{B}^{\rm target} = 2~002$ vibrational states.
This means we must include in the fully diagonalized block 1 all states with
$N_{V}^{(1)} \le 9$ which gives $M_{B}^{(1)} = 48~620$.
The total size of the basis corresponds to $N_V^{\rm max} = 14$ and $N_B^{\rm max} = 817~190$  functions.
Computing the elements of block 2, whose size is
$M_B^{(1)} \times M_B^{\rm max} = 48620 \times 817190$, and including them as
a perturbation takes 3 hours on an 8-core desktop computer. This is cheap compared
to the subsequent diagonalization of the 48~620 dimensional matrix.

\begin{table}[ht]
\scriptsize
\tabcolsep=3pt
\caption{Experimental and calculated vibrational term values, in \cm, for HNO\3.
Calculation I is based on the {\it ab initio} parameters while Calculations II and III
are the results of the use of the refined PES values.
The references give the source of the experimental data used, but see text for a discussion
of the actual values given.} \label{tab:freq}
\begin{center}
%\resizebox{\linewidth}{!}{%
\begin{tabular}{lcrrrrc}
\hline\hline
\multicolumn{2}{c}{State} & I &  II & III &  Exp. & Source\\
\hline
$A^{"}$  &  \n\9        &   456.7  &   458.2  &   458.2  &   458.2  &  \cite{94PeFlCA.HNO3}  \\
$A^{'}$  &  \n\8        &   577.0  &   580.3  &   580.4  &   580.3  &  \cite{98Pexxxx.HNO3}  \\
$A^{'}$  &  \n\7        &   647.3  &   646.7  &   647.0  &   646.5  &  \cite{98Pexxxx.HNO3}  \\
$A^{"}$  &  \n\6        &   770.1  &   763.2  &   763.3  &   763.1  &  \cite{pnnl}           \\
$A^{'}$  &  \n\5        &   876.7  &   879.1  &   878.8  &   879.1  &  \cite{pnnl}           \\
$A^{'}$  & 2\n\9        &   894.1  &   896.4  &   896.2  &   896.3  &  \cite{pnnl}           \\
$A^{"}$  &  \n\8+\n\9   &  1025.0  &  1029.9  &  1038.0  &  1038.0  &  \cite{pnnl}           \\
$A^{"}$  &  \n\7+\n\9   &  1096.0  &  1097.0  &  1100.3  &  1100.8  &  \cite{pnnl}           \\
$A^{'}$  &  \n\6+\n\9   &  1210.2  &  1205.4  &  1205.2  &  1205.6  &  \cite{pnnl}           \\
$A^{"}$  & 3\n\9        &  1284.2  &  1288.8  &  1289.6  &  1289.0  &  \cite{94PeFlCA.HNO3}  \\
$A^{'}$  &  \n\4        &  1303.0  &  1302.9  &  1303.2  &  1303.1  &  \cite{pnnl}           \\
$A^{'}$  &  \n\3        &  1329.6  &  1326.2  &  1326.3  &  1325.7  &  \cite{pnnl}           \\
$A^{"}$  &  \n\5+\n\9   &  1337.5  &  1340.5  &  1343.7  &  1343.6  &  \cite{94PeFlCA.HNO3}  \\
$A^{'}$  &  \n\7+\n\5   &  1507.9  &  1509.9  &  1516.0  &  1515.9  &  \cite{pnnl}           \\
$A^{'}$  & 2\n\6        &  1539.2  &  1525.4  &  1525.4  &  1525.6  &  \cite{pnnl}           \\
$A^{'}$  &  \n\7+2\n\9  &  1526.1  &  1528.2  &  1533.7  &  1533.2  &  \cite{pnnl}           \\
$A^{'}$  & 4\n\9        &  1656.7  &  1662.8  &  1661.3  &  1664.7  &  \cite{94PeFlCA.HNO3}  \\
$A^{'}$  &  \n\2        &  1716.4  &  1709.5  &  1709.4  &  1709.6  &  \cite{pnnl}           \\
$A^{'}$  & 2\n\5        &  1747.1  &  1751.3  &  1757.0  &  1757.0  &  \cite{pnnl}           \\
$A^{'}$  &  \n\5+2\n\9  &  1769.9  &  1773.9  &  1780.4  &  1780.3  &  \cite{pnnl}           \\
$A^{"}$  &  \n\3+\n\9   &  1790.2  &  1788.0  &  1789.2  &  1789.7  &  \cite{pnnl}           \\
$A^{'}$  &  \n\3+\n\8   &  1900.6  &  1900.4  &  1905.8  &  1906.0  &  \cite{pnnl}           \\
$A^{'}$  &  \n\4+\n\7   &  1941.4  &  1940.8  &  1949.2  &  1949.6  &  \cite{pnnl}           \\
$A^{'}$  &  \n\3+\n\7   &  1968.4  &  1964.3  &  1974.7  &  1975.2  &  \cite{pnnl}           \\
$A^{"}$  &  \n\4+\n\6   &  2066.5  &  2059.7  &  2061.4  &  2061.4  &  \cite{pnnl}           \\
$A^{"}$  &  \n\3+\n\6   &  2102.4  &  2091.5  &  2091.9  &  2092.0  &  \cite{pnnl}           \\
$A^{"}$  &  \n\2+\n\9   &  2174.5  &  2169.7  &  2165.2  &  2164.8  &  \cite{pnnl}           \\
$A^{'}$  &  \n\2+\n\5   &  2537.5  &  2531.0  &  2530.8  &  2530.6  &  \cite{04FeHaVa.HNO3}  \\
$A^{'}$  & 2\n\4        &  2584.2  &  2582.1  &  2580.5  &  2580.9  &  \cite{04FeHaVa.HNO3}  \\
$A^{'}$  &  \n\2+2\n\9  &  2595.2  &  2593.2  &  2596.2  &  2596.5  &  \cite{04FeHaVa.HNO3}  \\
$A^{'}$  & 2\n\3        &  2651.0  &  2644.0  &  2643.8  &  2644.4  &  \cite{04FeHaVa.HNO3}  \\
$A^{'}$  &  \n\2+\n\4   &  3003.2  &  2998.2  &  2998.4  &  2998.5  &  \cite{04FeHaVa.HNO3}  \\
$A^{'}$  &  \n\2+\n\3   &  3033.3  &  3022.6  &  3021.8  &  3022.1  &  \cite{04FeHaVa.HNO3}  \\
$A^{'}$  & 2\n\2        &  3411.2  &  3396.7  &  3404.2  &  3404.4  &  \cite{04FeHaVa.HNO3}  \\
$A^{'}$  &  \n\1        &  3553.3  &  3551.6  &  3551.6  &  3551.9  &  \cite{pnnl}           \\
$A^{"}$  &  \n\1+\n\9   &  4007.3  &  4007.3  &  4006.6  &  4007.0  &  \cite{04FeHaVa.HNO3}  \\
$A^{'}$  &  \n\1+\n\8   &  4125.5  &  4127.3  &  4127.4  &  4127.5  &  \cite{04FeHaVa.HNO3}  \\
$A^{'}$  &  \n\1+\n\7   &  4199.4  &  4196.3  &  4196.8  &  4197.0  &  \cite{04FeHaVa.HNO3}  \\
$A^{'}$  & 2\n\2+2\n\9  &  4319.1  &  4314.0  &  4315.0  &  4314.5  &  \cite{04FeHaVa.HNO3}  \\
$A^{'}$  &  \n\1+\n\5   &  4427.6  &  4428.2  &  4427.4  &  4427.6  &  \cite{04FeHaVa.HNO3}  \\
$A^{'}$  &  \n\1+2\n\9  &  4445.5  &  4446.0  &  4445.5  &  4445.8  &  \cite{04FeHaVa.HNO3}  \\
$A^{'}$  & 2\n\2+\n\3   &  4757.6  &  4751.5  &  4751.9  &  4750.0  &  \cite{04FeHaVa.HNO3}  \\
$A^{"}$  &  \n\1+3\n\9  &  4831.0  &  4833.4  &  4833.9  &  4832.8  &  \cite{04FeHaVa.HNO3}  \\
$A^{'}$  &  \n\1+\n\4   &  4870.7  &  4866.8  &  4865.5  &  4866.3  &  \cite{04FeHaVa.HNO3}  \\
$A^{'}$  &  \n\1+\n\2   &  5256.6  &  5248.2  &  5254.5  &  5252.4  &  \cite{04FeHaVa.HNO3}  \\
$A^{'}$  & 2\n\1        &  6941.5  &  6935.1  &  6938.8  &  6940.0  &  \cite{04FeHaVa.HNO3}  \\
\hline\hline
\end{tabular}
%}
\end{center}
\end{table}

Table~\ref{tab:freq} shows the calculated and experimental vibrational term
values for the HNO\3 molecule. These calculations were carried out with our
initial, {\it ab initio}, PES.  As can be seen, this PES gives a generally
satisfactory description of the experimental vibrational term values. The
average deviation between the calculated and experimental fundamental energy
levels is 3~\cm. For the first overtone and combination levels it is 6~\cm. The
calculation describes the strong Fermi resonance between \n\5 and 2\n\9 well.

However, this accuracy is not sufficient to generate a good line list.
Therefore, we have refined the PES parameters using the method of
regularization.  The parameters $\alpha_i$, which define the
half-width of the Morse functions for the stretching coordinates, were
fixed in the fits to their {\it ab initio} values. The inverse problem
was solved in two steps. First the 39 quadratic parameters of the potential
function, $D_{ij}$ of Eq.~(\ref{e:potential}), were refined using the
11 equally-weighted vibrational term values: the fundamentals plus
2\n\9 and \n\6+\n\9.  The results of this fit are given as calculation
II in Table~\ref{tab:freq}.  The average deviation between calculated
and experimental fundamental levels energy is now less than 0.2~\cm\
and is about 3~\cm\ for the first overtone and first combination
bands.

In the second stage all potential parameters $D_{ij}$ , $D_{ijk}$ and
$D_{ijkl}$ are processed: a total of 584 parameters were refined using the  46
experimental term values, given in Table~\ref{tab:freq}. Following the
regularization method we use 584 additional constraints to for these parameters
to their initial {\it ab initio} $D_{ijk}$ and $D_{ijkl}$ values  and to the values
of $D_{ij}$ obtained at the previous stage. This makes the inverse problem fully
determined despite the small amount of experimental data. States with up
to 3 quanta of excitation had weight 1.0, while 4 quanta states, whose energies
are more uncertain, had a weight of 0.1. The results of this fit are shown as
calculation III in Table~\ref{tab:freq}. The average deviation between
calculated and experimental fundamental levels energy remains 0.2~\cm, but for
the first overtone and first combination bands it is reduced to 0.4~\cm. This
potential function was subsequently used to calculate the ro-vibrational energy
levels and the line list.

We note  that the `experimental' values of the vibrational term values given in
Table~\ref{tab:freq} do not match those given by Perrin {\it et al}
\cite{94PeFlCA.HNO3} or Feierabend {\it et al}  \cite{04FeHaVa.HNO3}. This is
because in these laboratory studies the corresponding vibrational term values were
estimated as band centers, either as maxima of the Q-branches or minima between
the P- and R-branches, and thus do not precisely correspond to the $J=0$ energy
of the vibrationally excited state, because they also contain some rotational
structure. Therefore, Table~\ref{tab:freq} gives our revised values:
vibrational term values were determined as the energy, which, after their
substitution in a full ro-vibrational calculation, gives coincidence between
the computed and experimental rotational structure of the absorption band in
question. These values should represent the best available estimate for the
HNO\3 vibrational term values. Our estimated accuracy for the `experimental'
vibrational term values given in Table~\ref{tab:freq} is better than 0.1~\cm.
The typical difference between our values and those given previously
\cite{94PeFlCA.HNO3, 04FeHaVa.HNO3} is about 1~\cm, which corresponds to the
half-width of a typical Q-branch. Finally, some experimental values given in
Table~\ref{tab:freq} do not come from high resolution spectra; these were
identified by us from the observed absorption cross sections provided by PNNL
\cite{pnnl}.

\section{Calculated ro-vibrational spectrum}
\label{s:Ro-vibrational}

We compute the rotation-vibration spectrum in the range  0--7000~\cm,
which includes all first overtones and first combination bands. To include all
hot absorption bands present at room temperature it was necessary to included
all vibrational states lying below 9000~\cm.

There are about 20~000 vibrational states below 9000~\cm.  Therefore, the
calculation of ro-vibrational energy levels for all of these vibrational states
and the calculation of intensities of allowed transtions between all the
ro-vibrational levels is lengthy, even when using our hybrid method. Initially,
this calculation took about 6 months on a 8-core desktop computer.  However
as described below, this time can be reduced by two orders-of-magnitude by computing only
those ro-vibrational energy levels and transition intensities which are actually needed.
For concreteness, in what follows we consider the explicit example of the
calculation of a room temperature spectrum.

First, room temperature spectrum experiments do not show any significant
transitions to vibrational states with the polyad number $N_{V} > 4$, due to
the very low intensity of such bands. Therefore, we only need to obtain results
for the vibrational states with the polyad number $N_{V}^{\rm   target} \le 4$.
This reduces the number of vibrational states for which ro-vibrational energy
levels are required to $M_{B}^{\rm target} = 1~715$. All other ro-vibrational
states are only used to perturb  the target ro-vibrational energy levels.

Second, not all of the 20~000 vibrational states below 9000~\cm\ actually
significantly contribute the target ro-vibrational energy levels. This is
because a large difference between the quantum numbers $v^{\lambda}_i$ and
$v^{\lambda\p}_i$ from vibrational states $\lambda$ and $\lambda\p$ leads to
vanishingly small values of the corresponding matrix elements
$\bar{\mu}^{\lambda \lambda\p}_{ab}$, Eq.~(\ref{e:M-matrix}), and
$H^{J}_{\lambda k, \lambda\p k\p}$, Eq.~(\ref{e:H-matrix}). For example, for
purely harmonic basis functions, the matrix elements  $\bar{\mu}^{\lambda
\lambda\p}_{ab}$ and $H^{J}_{\lambda k, \lambda\p k\p}$ are exactly zero  for
$\sum_i |v^{\lambda}_i - v^{\lambda\p}_i| > 2$.  Using a mixed Morse-harmonic
basis we obtain $\bar{\mu}^{\lambda \lambda\p}_{ab} \approx 0$ and
$H^{J}_{\lambda k, \lambda\p k\p} \approx 0$ for $\sum_i |v^{\lambda}_i -
v^{\lambda\p}_i| > 3$. Therefore, only the contribution from the vibrational
states with polyad numbers $N_V^{\rm vib} \le 7$ need to be evaluated as
perturbation to the target ro-vibrational energy levels from the $M_B^{\rm
vib}$ vibrational states. 
In this case, $M_B^{\rm vib} = 9~477$, which is only about half the vibrational states
below 9000~\cm. Besides, when summing the perturbation effect
for a given ro-vibrational energy levels, we can skip all pairs with $\sum_i
|v^{\lambda}_i - v^{\lambda\p}_i| > 3$. Therefore, the sums in
Eqs.~(\ref{e:H-diag}) and (\ref{e:H-nondiag}) for each value of $\lambda$ will
run over less than a tenth of all the levels included in $M_B^{\rm vib}$.

Third, when considering a transition between different ro-vibrational
states, it is useful to make a preliminary assessment of its
intensity.  If the estimated value is below some threshold, the
intensity calculation can be skipped.  Such intensities can be neglected
either because of the small intrinsic value of the transition
dipole or because of the low population of the initial ro-vibrational
energy level caused by the Boltzmann factor.  As, in large line
lists, computation of the transition intensities dominates the
computer time \cite{jt564}, this significantly reduces the overall
computer time.

When these three factors are taken into account, the time for computing the
ro-vibrational spectrum in the 0 -- 7000~\cm\ region is reduced from 6 months
to two days on an 8-core desktop computer. This is quick enough even to allow
us to refine our {\it ab initio} DMF by fitting to experimental line
intensities, thus improving agreement between observed and computed spectra.

In this case, the DMF parameters $\vec D^{0}$ , $\vec d_{i}$ and $\vec d_{ij}$
of Eq.~(\ref{e:dipol}) were varied to achieve the best agreement between
theoretical and experimental integrated transition intensities for a given
spectral region. This fit was again conducted using the method of
regularization. We used experimental cross sections from the PNNL databse
\cite{pnnl} as input data. In the low-frequency region, which is absent from
the PNNL spectrum, we used  data from  HITRAN \cite{jt557}. In the
high-frequency region 4200 -7000 \cm, where the PNNL spectra are very noisy, we
used the experimental intensities of Feierabend {\it et al}
\cite{04FeHaVa.HNO3}. It should be noted that Feierabend {\it et al} give only
relative intensities, where the \n\1 band intensity was taken as unity. For
this region, Table~\ref{tab:int} gives absolute intensities obtained by
multiplying the relative intensities of Feierabend {\it et al} by the absolute
intensity of the \n\1  band from PNNL. Absolute intensity values for PNNL
\cite{pnnl} were obtained by integrating the PNNL absorption cross sections.
Whenever possible, we sought not only to have the best agreement between the
calculated and experimental integral intensities for a given spectral range,
but also good agreement between the intensities of individual transitions
within each of the spectral bands. In the case of complex absorption bands,
which are formed from the superposition of several intense bands, we used those
intensities which gave the best fit between the experimental and calculated
shape of the absorption band.

\begin{table}[ht]
\scriptsize
\tabcolsep=3pt
\caption{Experimental and calculated intensities by region for HNO\3: experimental
data is taken from PNNL \cite{pnnl} , HITRAN \cite{jt557} and Feierabend {\it et al} \cite{04FeHaVa.HNO3}.
Calculation A used an {\it ab initio} DMF and calculation B a fitted DMF.
The dominant bands for each frequency window are also given.} \label{tab:int}
\begin{center}
%\resizebox{\linewidth}{!}{%
\begin{tabular}{lrrrrrr}
\hline\hline
Band& Frequency (\cm) & \multicolumn{4}{c}{Intensity (km/mole)} \\
\cline{3-7}
                                               &   & \cite{jt557} & \cite{pnnl} & \cite{04FeHaVa.HNO3} & Calc. A & Calc. B \\
\hline
Rotation                                       &   0   -  100  &    6.7  &         &         &    7.2   &    6.7   \\
Hot                                            &   100 -  350  &         &         &         &    0.15  &    0.14  \\
\n\9                                           &   350 -  520  &   77.7  &         &         &  107.3   &   77.7   \\
\n\8                                           &   520 -  610  &    5.7  &         &         &    6.4   &    5.6   \\
\n\7                                           &   610 -  700  &    5.6  &         &         &   13.2   &    5.6   \\
\n\6                                           &   700 -  830  &    7.4  &         &         &    6.9   &    7.4   \\
\n\5 , 2\n\9                                   &   830 -  950  &  124.5  &  110.9  &         &  151.4   &  110.9   \\
\n\8+\n\9 , \n\7+\n\9                          &   950 - 1140  &         &         &         &    0.53  &    0.82  \\
\n\6+\n\9                                      &  1140 - 1240  &    5.7  &    7.9  &         &    8.7   &    7.9   \\
\n\4 , \n\3 , 3\n\9 , \n\5+\n\9                &  1240 - 1380  &  229.3  &  221.6  &         &  300.4   &  220.8   \\
\n\7+\n\5 , 2\n\6 , \n\7+2\n\9                 &  1380 - 1600  &         &    6.5  &         &    4.7   &    6.1   \\
\n\2 , 4\n\9 , 2\n\5 , \n\5+2\n\9 , \n\3+\n\9  &  1600 - 1825  &  263.5  &  251.8  &         &  354.0   &  251.4   \\
\n\3+\n\8 , \n\4+\n\7 , \n\3+\n\7              &  1825 - 2040  &         &    1.5  &         &    3.4   &    1.8   \\
\n\3+\n\5 , \n\2+\n\9                          &  2170 - 2240  &         &   0.37  &   0.44  &    0.67  &    0.38  \\
\n\2+\n\5 , 2\n\4 , \n\2+2\n\9 , 2\n\3         &  2460 - 2710  &         &   6.2   &   8.2   &    8.4   &    6.1   \\
\n\2+\n\4 , \n\2+\n\3                          &  2920 - 3055  &         &   6.6   &   8.2   &    3.6   &    6.4   \\
2\n\2                                          &  3360 - 3440  &         &   1.3   &   1.6   &    0.51  &    1.3   \\
\n\1                                           &  3490 - 3610  &         &   54.9  &   54.9  &   76.9   &   54.9   \\
3\n\4 , \n\3+2\n\4                             &  3828 - 3893  &         &   0.08  &   0.11  &    0.33  &    0.34  \\
\n\1+\n\9                                      &  3950 - 4050  &         &   0.98  &   1.1   &    1.9   &    1.0   \\
\n\1+\n\8                                      &  4075 - 4160  &         &   0.19  &   0.27  &    0.53  &    0.35  \\
2\n\2+2\n\9                                    &  4230 - 4355  &         &   0.12  &   0.22  &    0.61  &    0.55  \\
\n\1+\n\5 , \n\1+2\n\9                         &  4385 - 4490  &         &   0.13  &   0.16  &    0.20  &    0.18  \\
2\n\2+\n\3 , 2\n\2+\n\4                        &  4630 - 4710  &         &   0.06  &   0.27  &    0.62  &    0.48  \\
2\n\2+\n\3                                     &  4710 - 4780  &         &   0.13  &   0.11  &    0.67  &    0.24  \\
\n\1+\n\4 , \n\1+\n\3  , \n\1+3\n\9            &  4790 - 4905  &         &   0.89  &   1.0   &    1.7   &    1.1   \\
3\n\2                                          &  5040 - 5115  &         &   0.06  &   0.05  &    0.59  &    0.50  \\
\n\1+\n\2                                      &  5210 - 5290  &         &   0.33  &   0.33  &    0.52  &    0.40  \\
\n\1+2\n\4                                     &  6080 - 6195  &         &   0.12  &   0.05  &    0.85  &    0.65  \\
2\n\1 , \n\1+2\n\2                             &  6865 - 7005  &         &         &   2.1   &    3.0   &    2.1   \\
\hline\hline
\end{tabular}
%}
\end{center}
\end{table}

Table~\ref{tab:int} shows experimental and calculated intensities for different
frequency regions using the initial (calculation A) and fitted (calculation B)
values of the parameters $\vec D^{0}$ , $\vec d_{i}$ and $\vec d_{ij}$ in the
DMF. This table shows that the use of the {\it ab initio} DMF leads to a
systematic overestimation of the calculated intensities: by an average of 40\%\
for the fundamentals bands and by 90\%\ for the first overtones and first
combination bands.  Fitting gives greatly improved agreement between the
calculated and experimental intensities.  In this case, the average difference
for the intensities of the fundamental transitions is only 0.3\%\ and for the
first overtone and combintation bands it is 40\%. These differences between the
computed and measured band intensities are within the experimental
uncertainties. For example in the region of the fundamental bands, intensities
from HITRAN are on average 10\%\ higher than the absorption cross sections
given by PNNL. At the same time, the intensity of the \n\6+\n\9 combination
band is 30\% less in HITRAN than PNNL.  In addition, the PNNL spectrum which we
used becomes very noisy for low intensity absorptions. Therefore, at present,
it does not make sense to further improve the agreement between calculated and
experimental intensities.

\begin{figure}[t!]
\centering
\includegraphics[width = 0.7\textwidth]{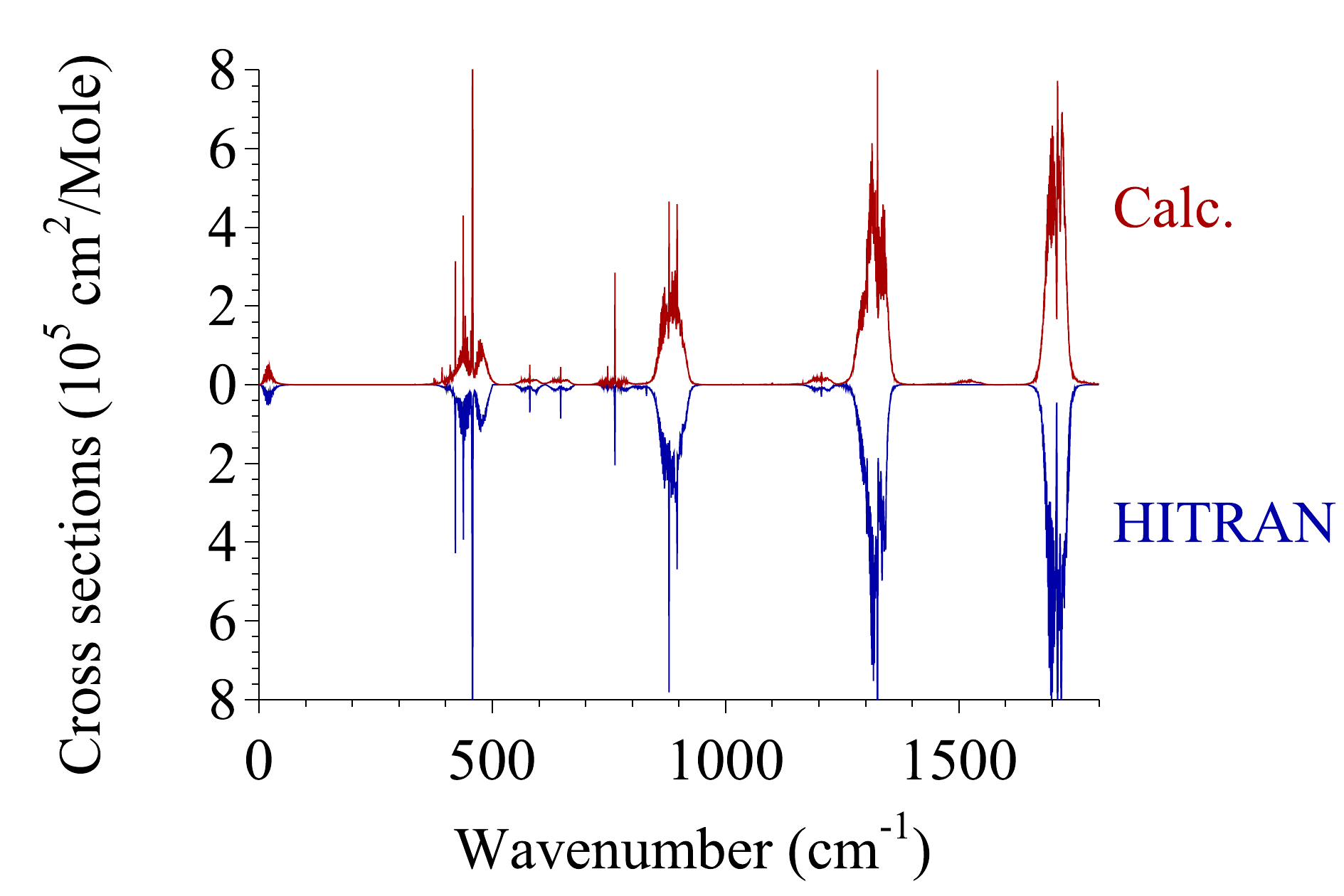}
\caption{Calculated (red curve) and HITRAN (blue curve) 296~K HNO\3 spectra in the
0 - 1800 \cm\ region.}
\label{fig:h1}
\end{figure}

\begin{figure}[t!]
\centering
\includegraphics[width = 0.7\textwidth]{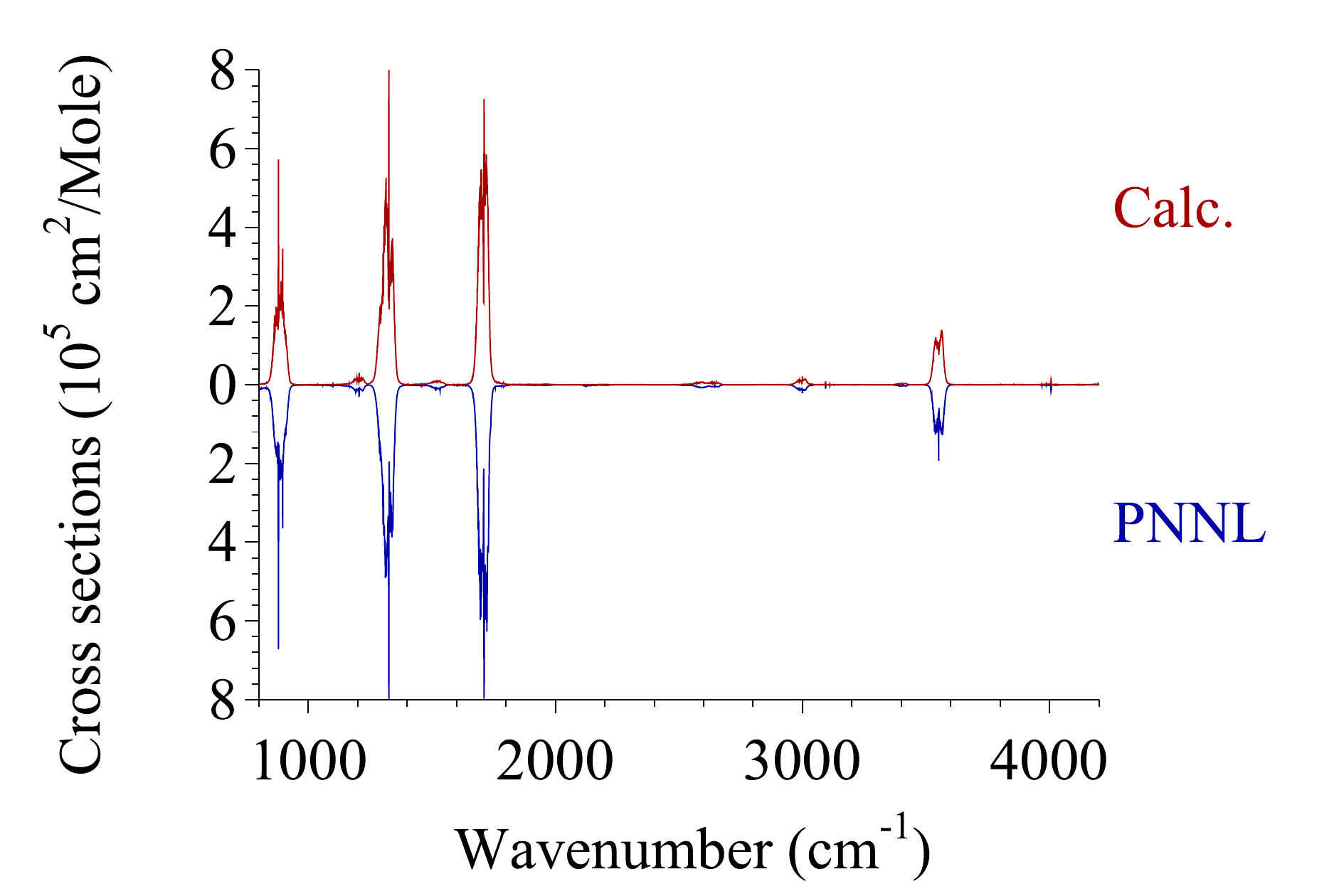}
\caption{Calculated (red curve) and experimental PNNL (blue curve) 298~K cross sections
in the 800 - 4200 \cm\ region.}
\label{fig:p1}
\end{figure}

Figure \ref{fig:h1}  compares our calculated spectra for HNO\3 at 296 K with
the data from HITRAN.  Although HITRAN aims to be comprehensive for
atmospherically important molecules such as HNO\3, it actually contains only a
few HNO\3 vibrational bands which means that HITRAN gives much less complete
coverage than the measured cross sections from PNNL. In particular, HITRAN has
no data for wavenumbers higher than 1900 \cm. Figures \ref{fig:p1} give a
similar overview comparison of our calculated spectrum with the 298~K PNNL
cross sections.

\begin{figure}[t!]
\centering
\includegraphics[width = 0.4\textwidth]{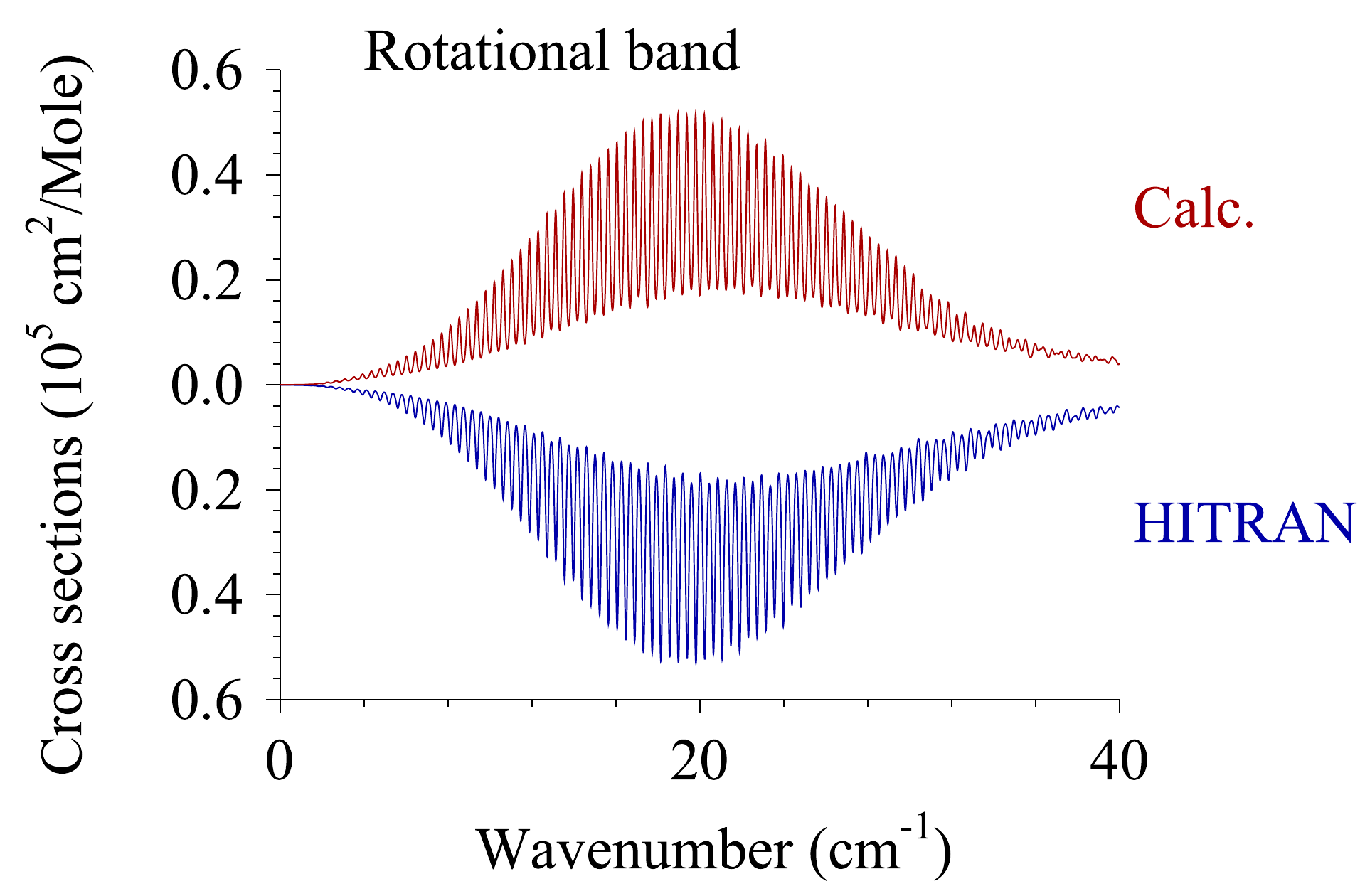}
\includegraphics[width = 0.4\textwidth]{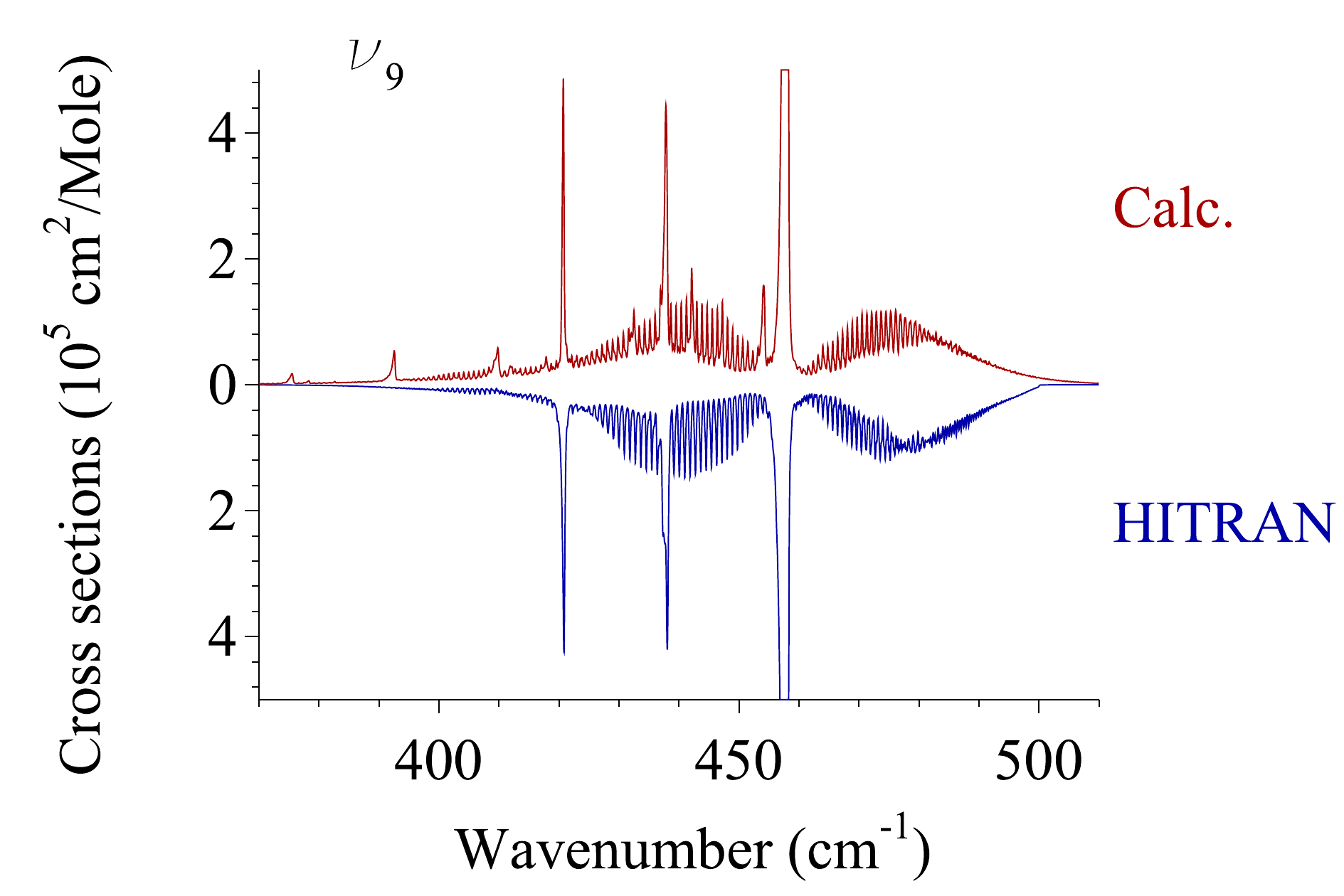}
\includegraphics[width = 0.4\textwidth]{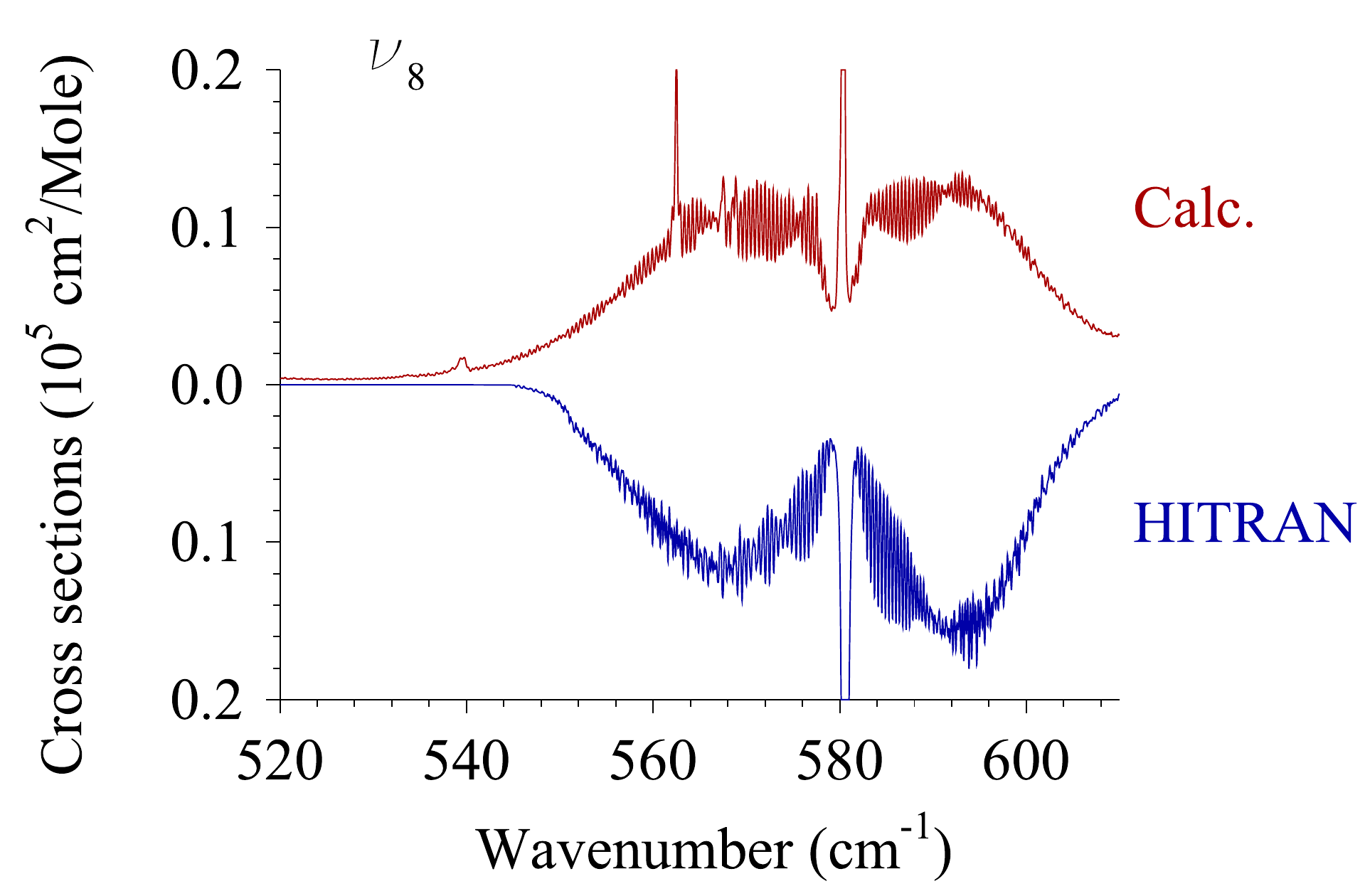}
\includegraphics[width = 0.4\textwidth]{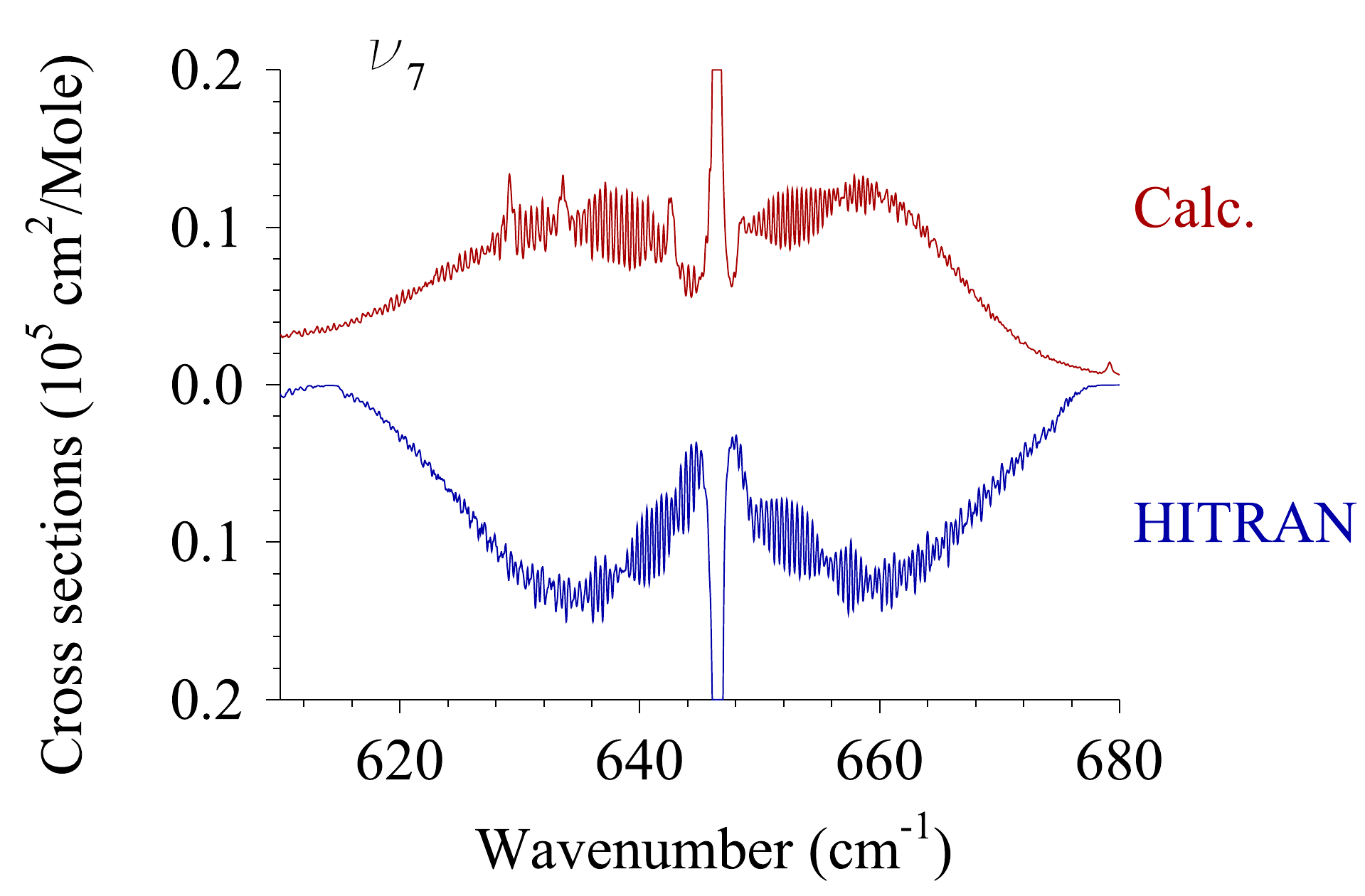}
\caption{Comparison of the main bands in HITRAN below 700 \cm: calculated (red curve) and HITRAN (blue curve)}
\label{fig:h2}
\end{figure}

\begin{figure}[t!]
\centering
\includegraphics[width = 0.4\textwidth]{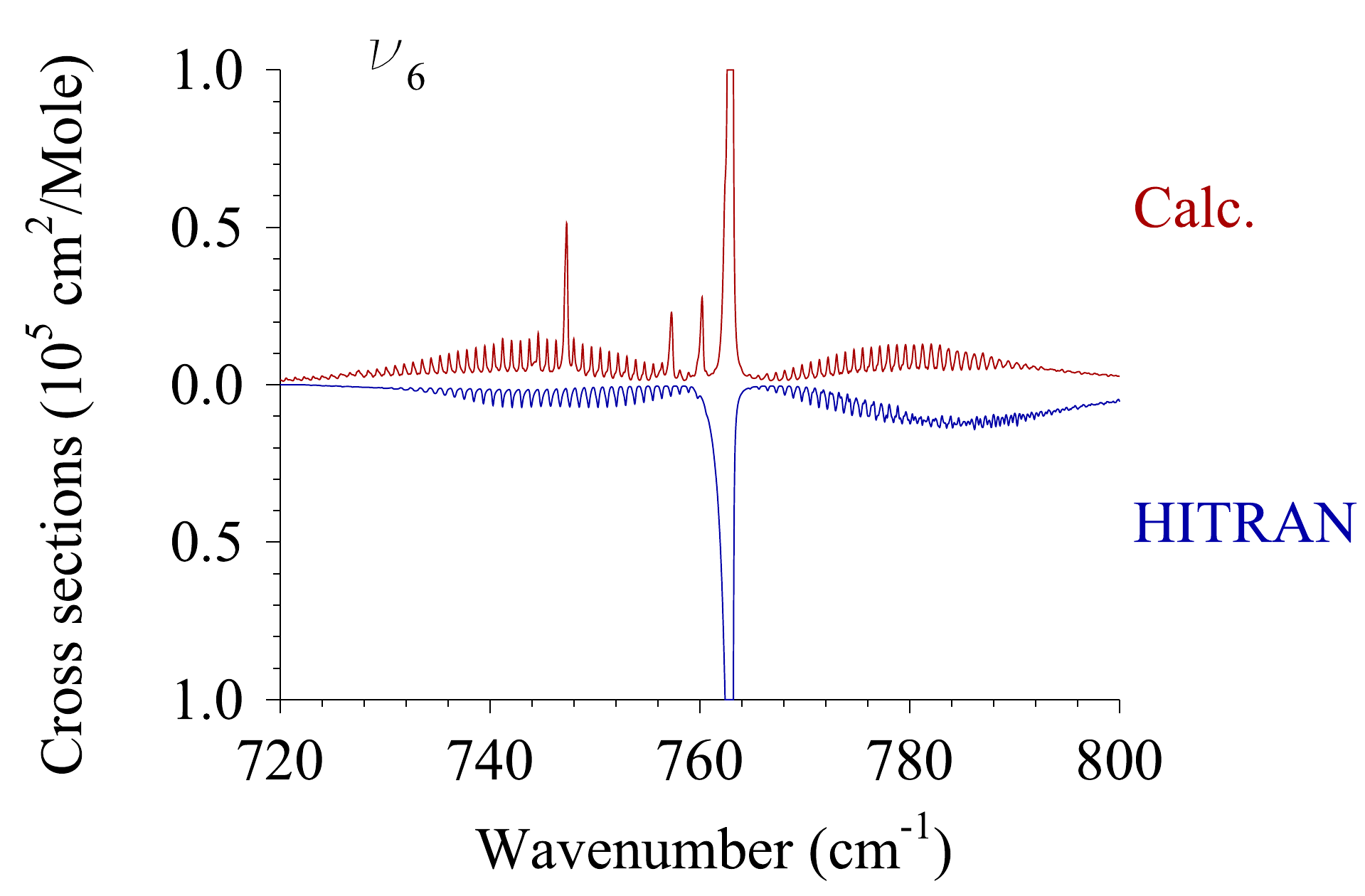}
\includegraphics[width = 0.4\textwidth]{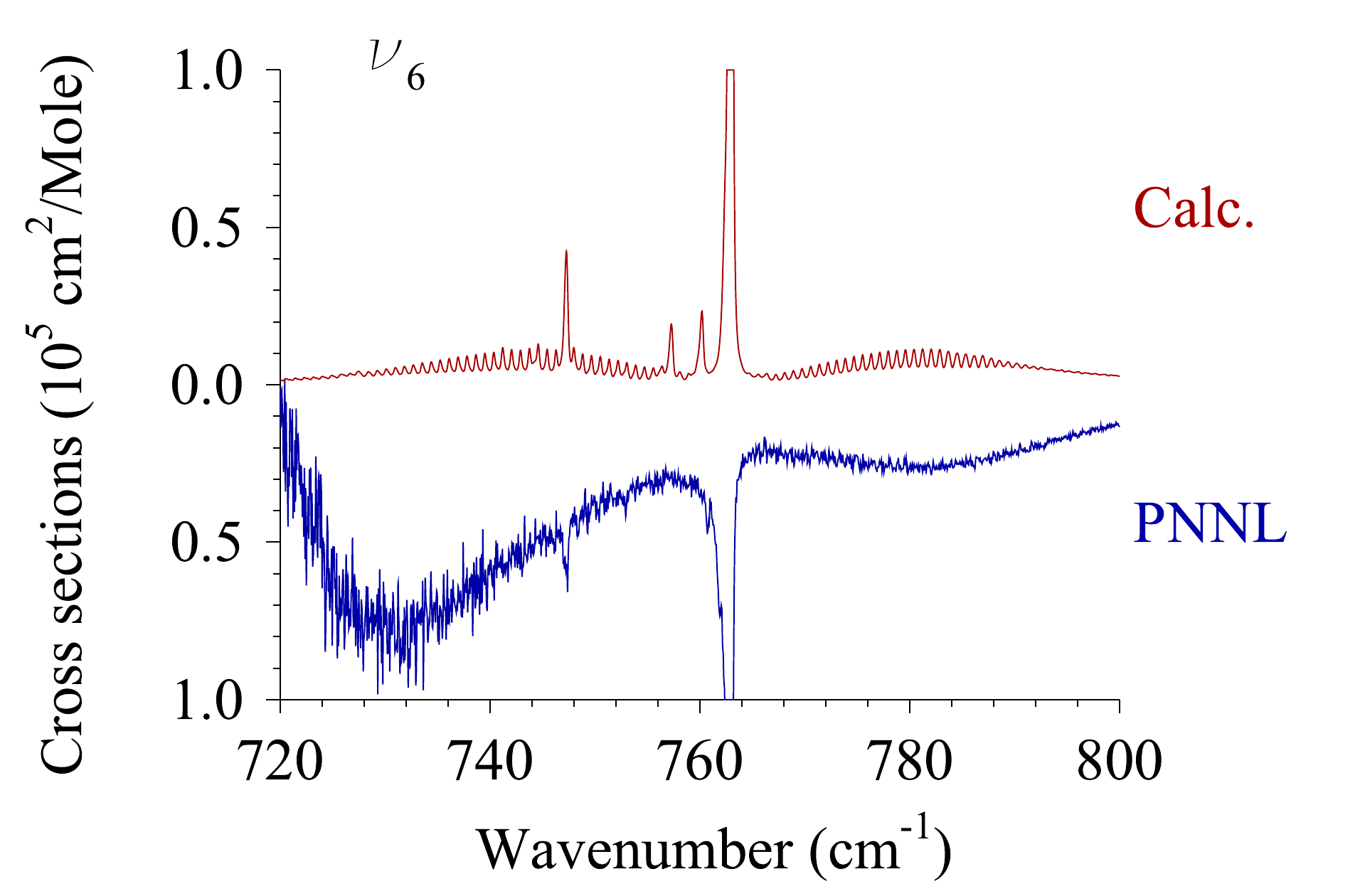}
\caption{Comparison of spectra for \n\6 band regions: Calculated (red curve) and experimental HITRAN/PNNL (blue curve).}
\label{fig:hp}
\end{figure}

Figure~\ref{fig:h2} presents more detailed comparisons for the main bands in
HITRAN below 700 \cm. Generally the agreement is very good.  HITRAN is
systematically missing data on hot bands even when they give rise to strong,
sharp features. For example, the \n\9 band region is missing several, intense
hot bands which are found in our calculation and which have been experimentally
observed \cite{94PeFlCA.HNO3}. A similar situation arises for the \n\6 band,
see Fig.~\ref{fig:hp}. This band is the only one for which a direct comparison
of HITRAN and PNNL data is possible. Again our calculations predict sharp
hot-band features which are absent from the HITRAN spectra. Despite becoming
increasingly noisy at low frequencies, the strongest of these hot-band features
can clearly be seen in the PNNL cross sections. It should be noted that the anharmonic
character of the torsion \n\9 mode gives rise to a sequence of hot bands in
the region 370--510~\cm, significantly shifted from the center of \n\9. 

\begin{figure}[ht!]
\centering
\includegraphics[width = 0.4\textwidth]{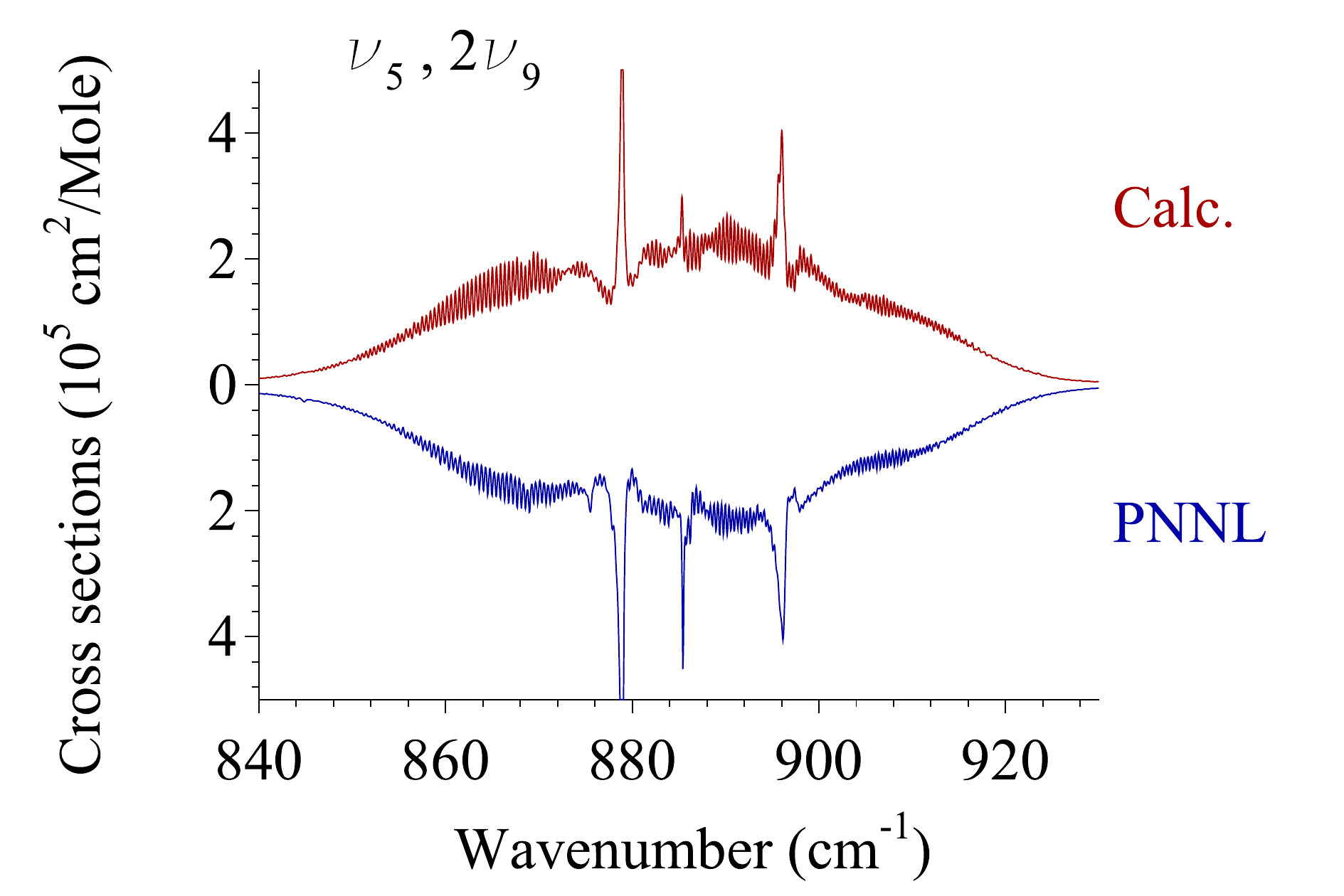}
\includegraphics[width = 0.4\textwidth]{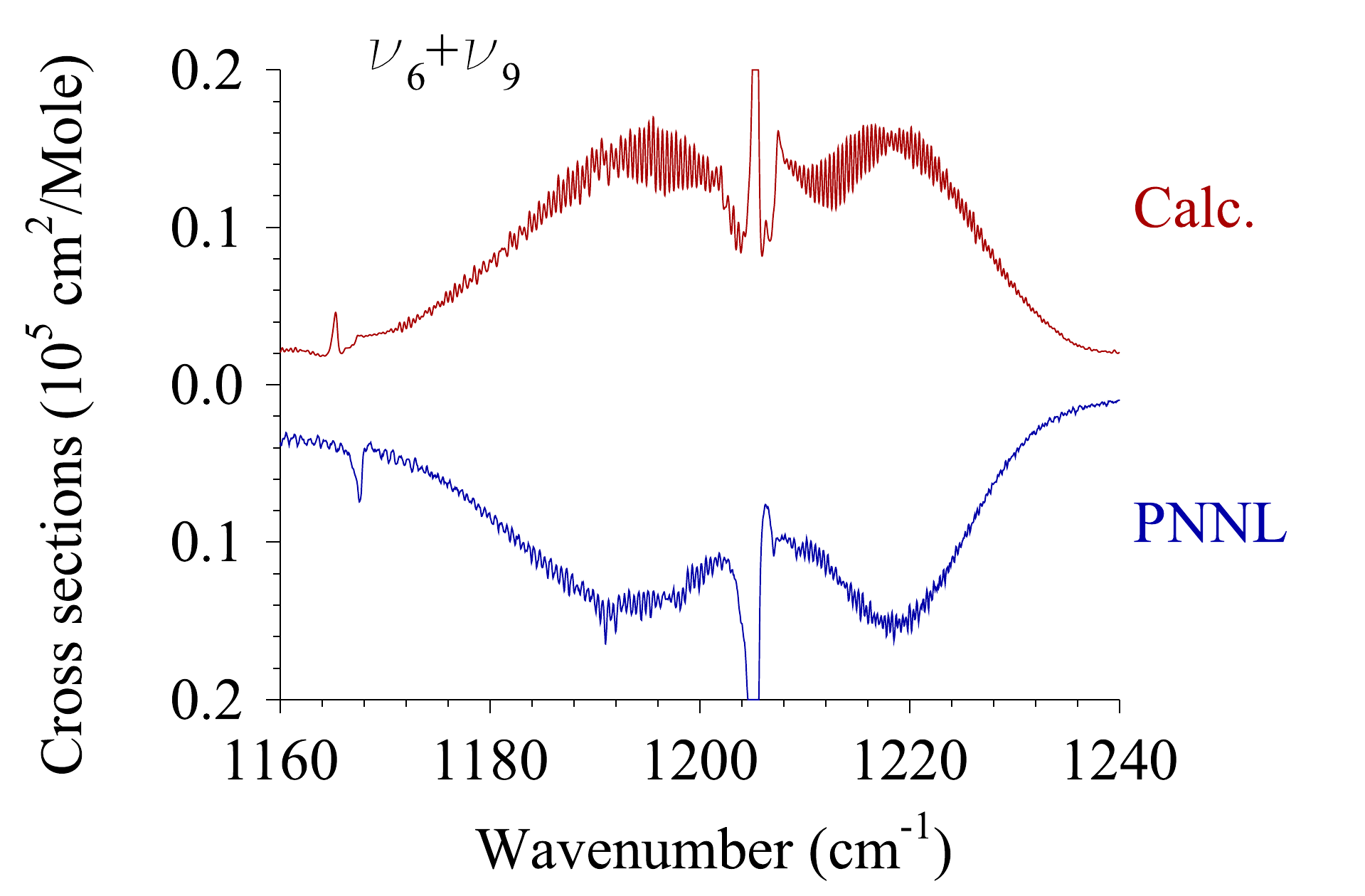}
\includegraphics[width = 0.4\textwidth]{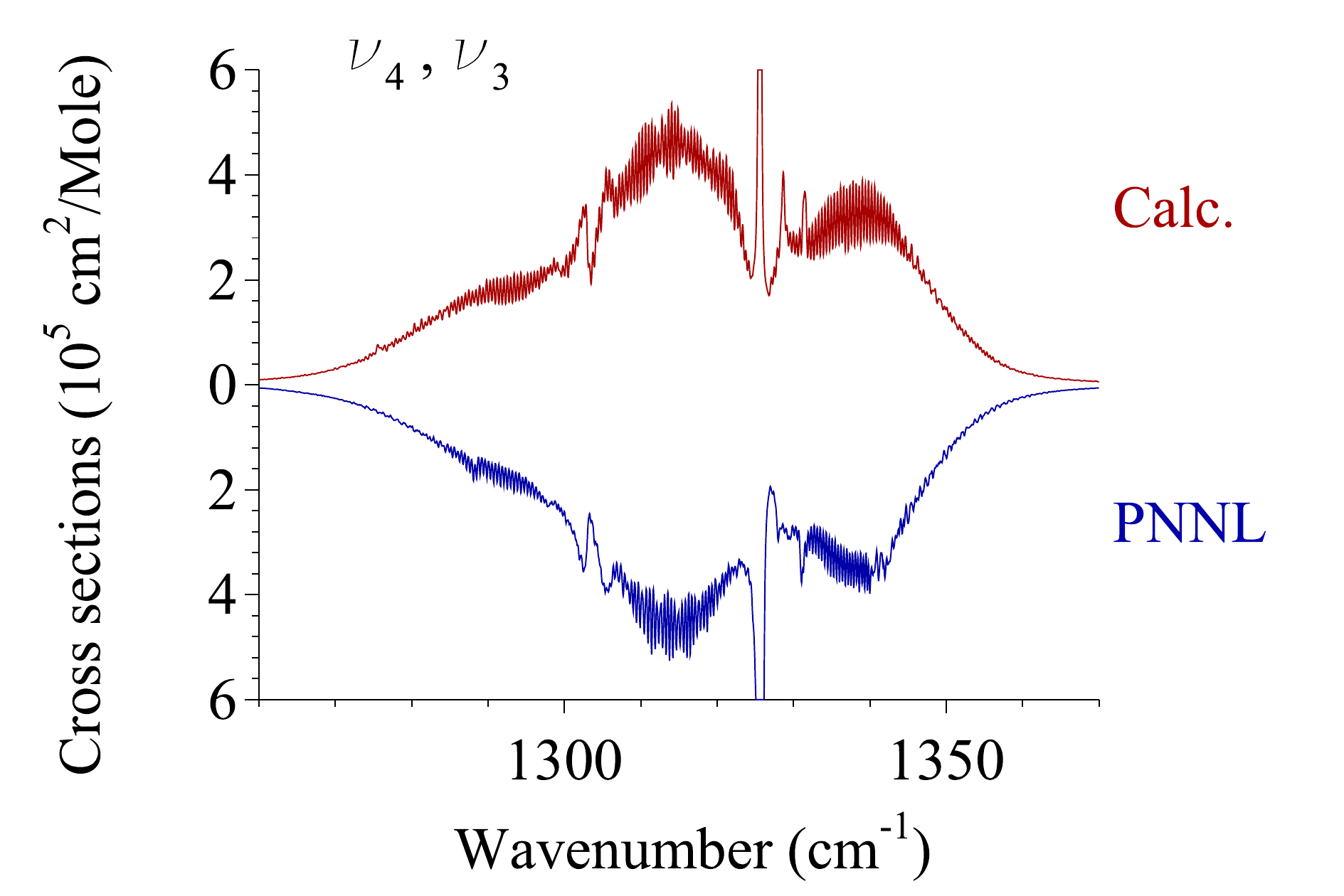}
\includegraphics[width = 0.4\textwidth]{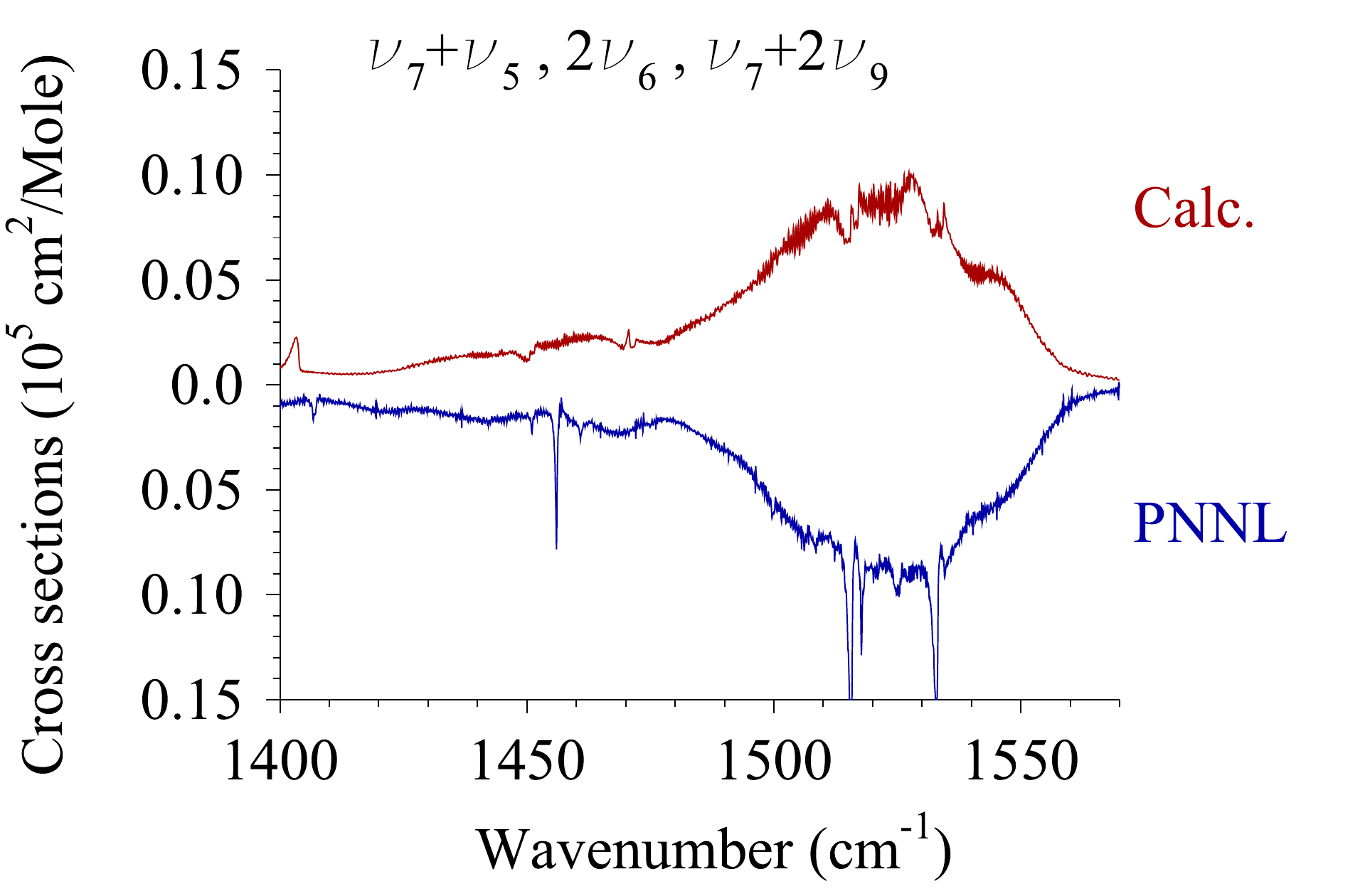}
\includegraphics[width = 0.4\textwidth]{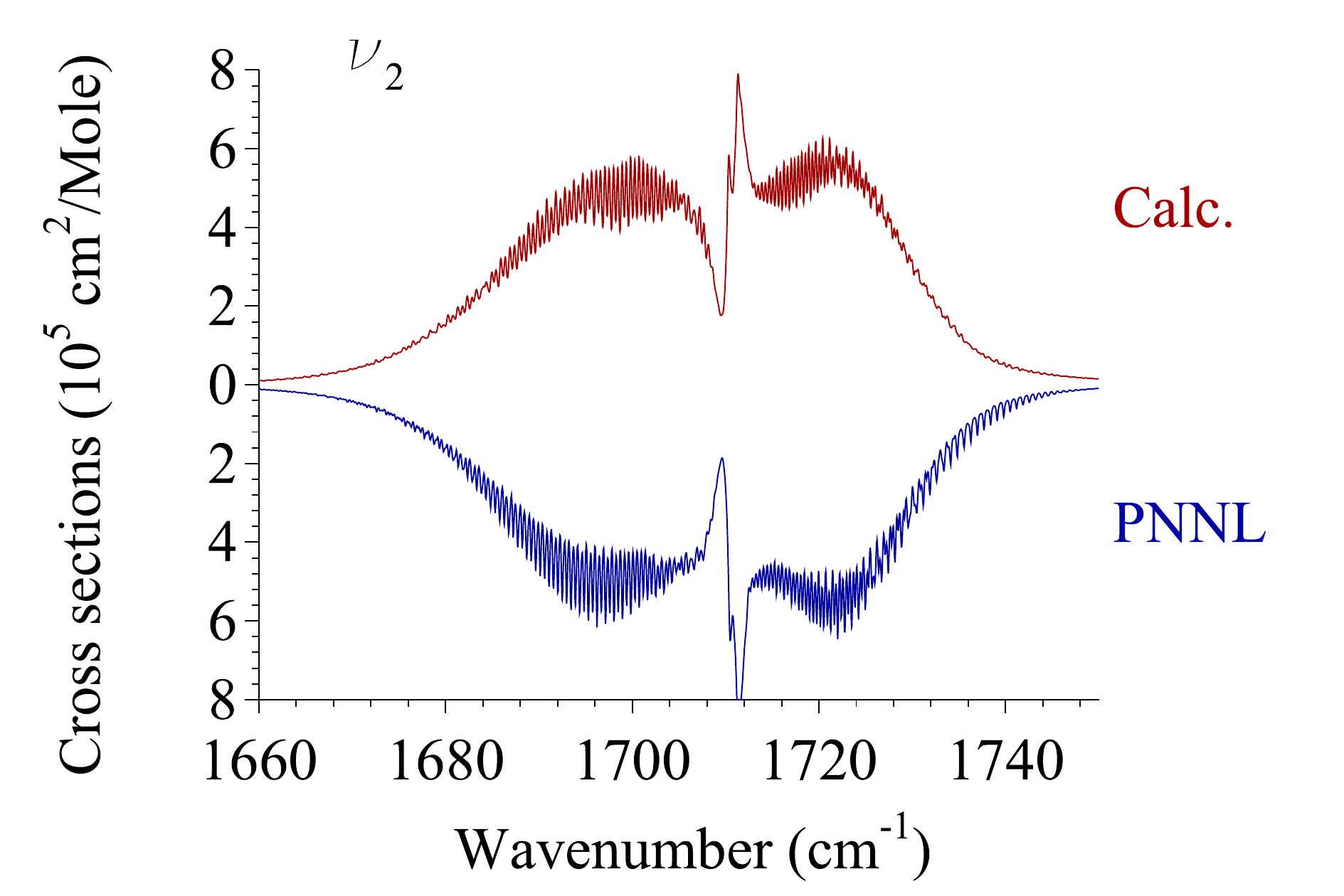}
\includegraphics[width = 0.4\textwidth]{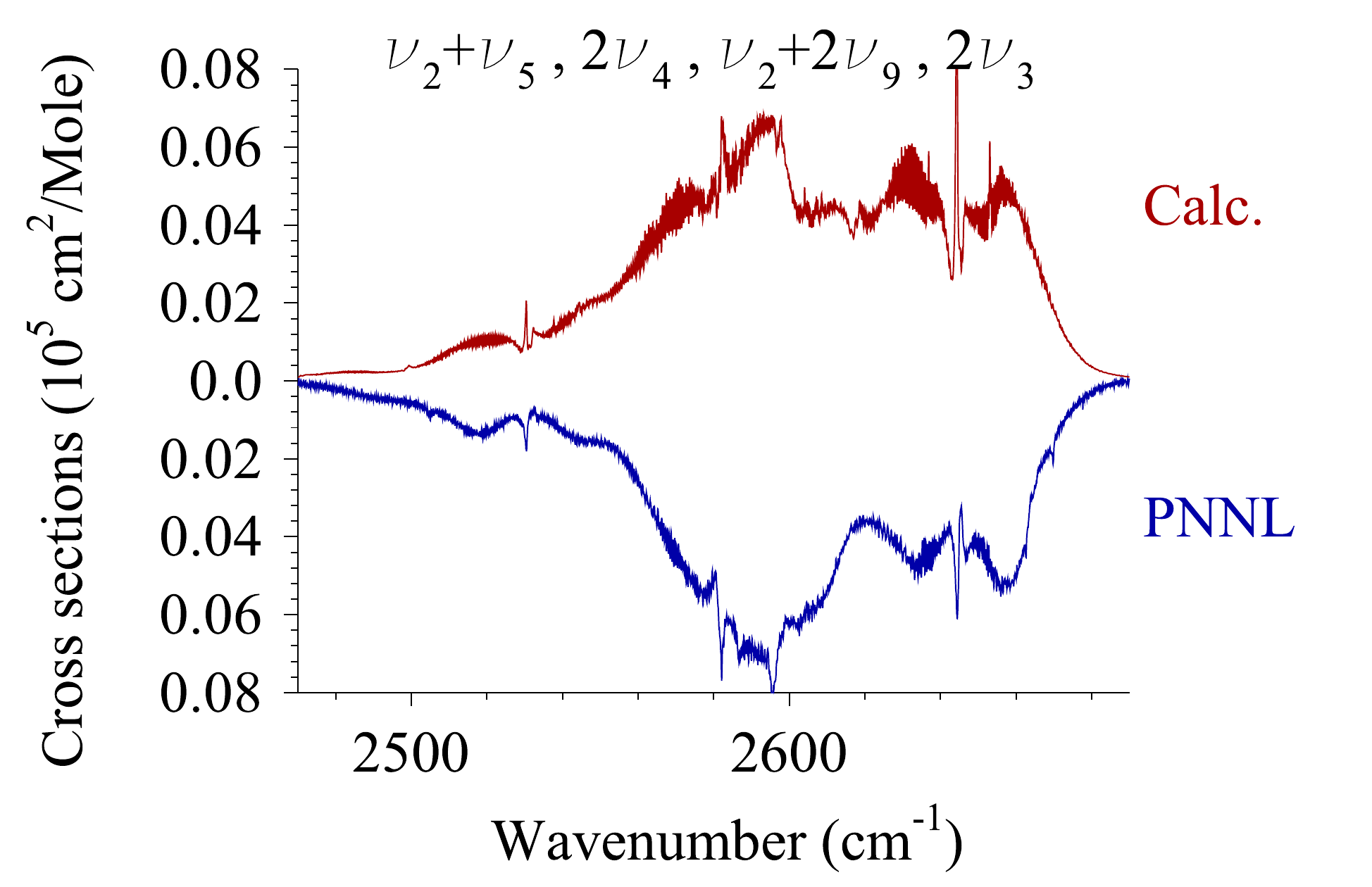}
\includegraphics[width = 0.4\textwidth]{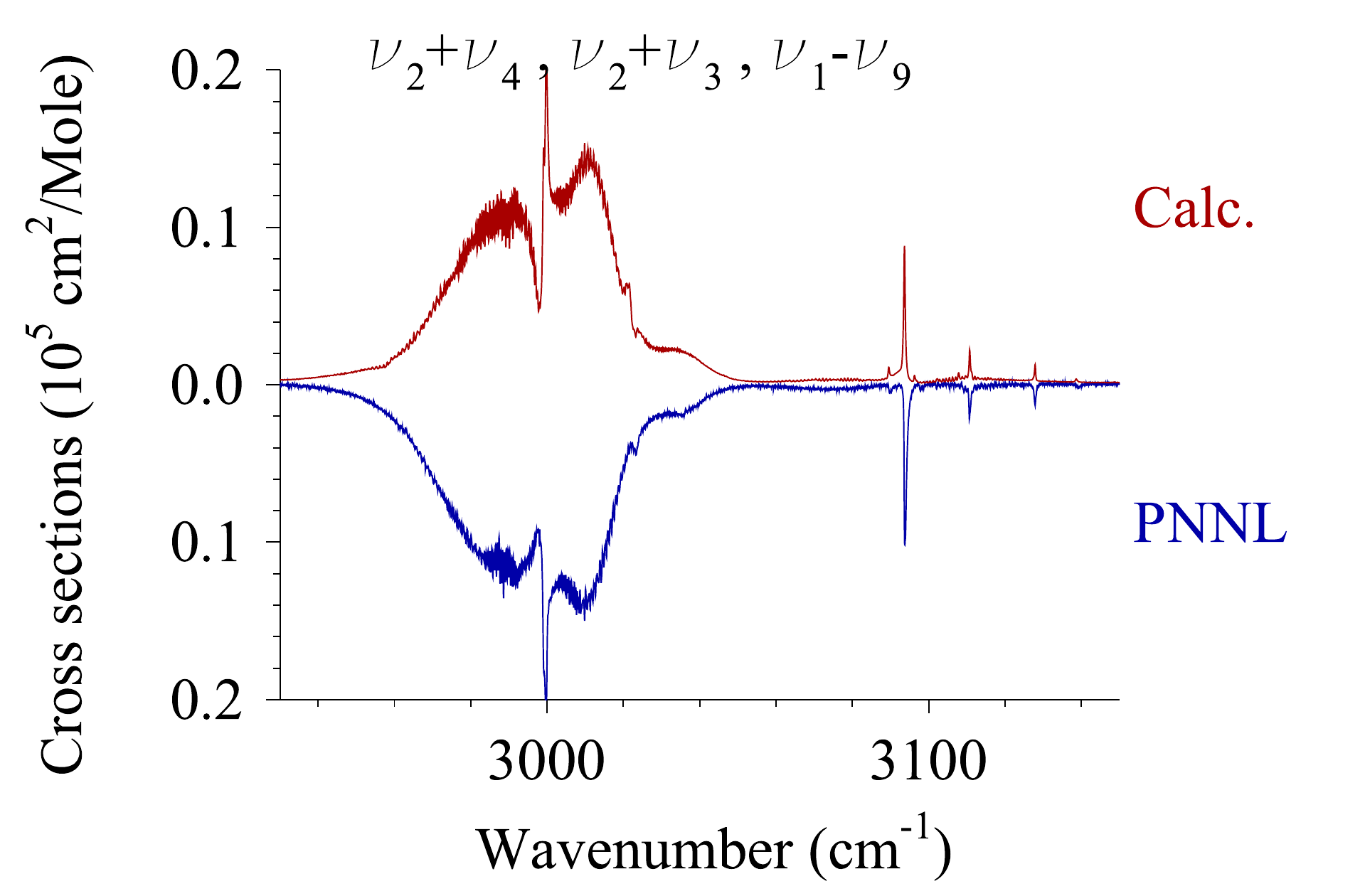}
\includegraphics[width = 0.4\textwidth]{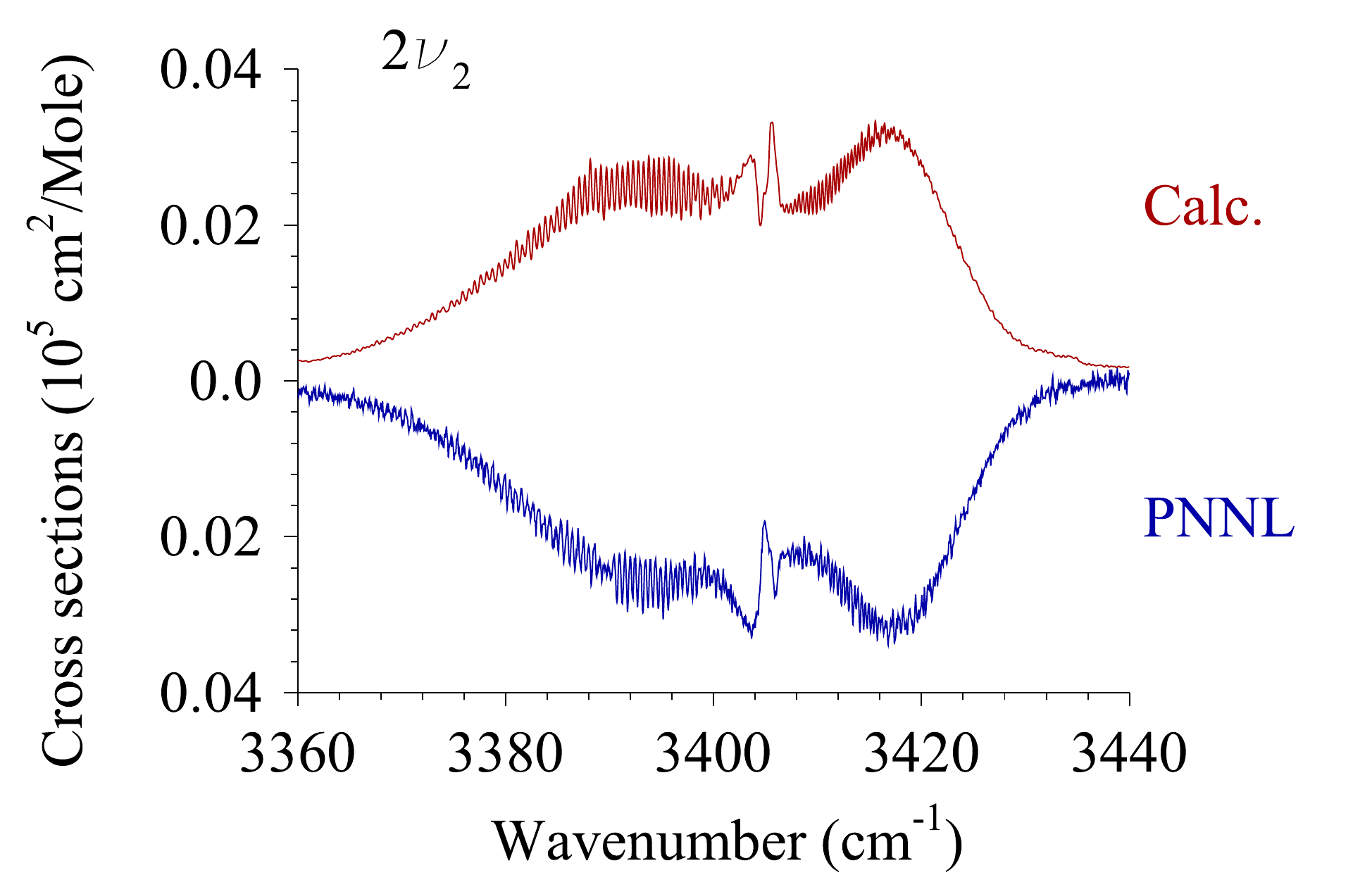}
\includegraphics[width = 0.4\textwidth]{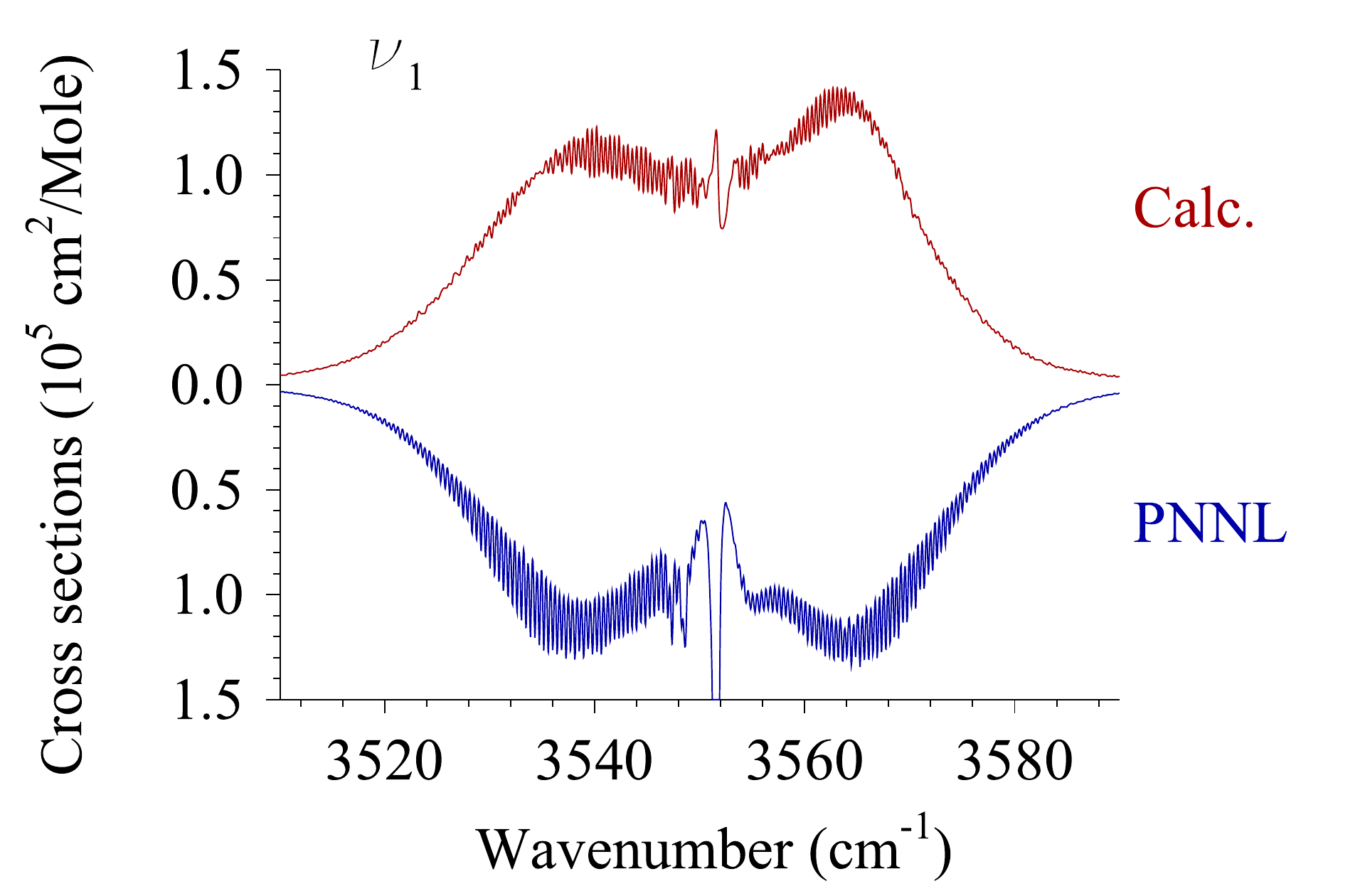}
\includegraphics[width = 0.4\textwidth]{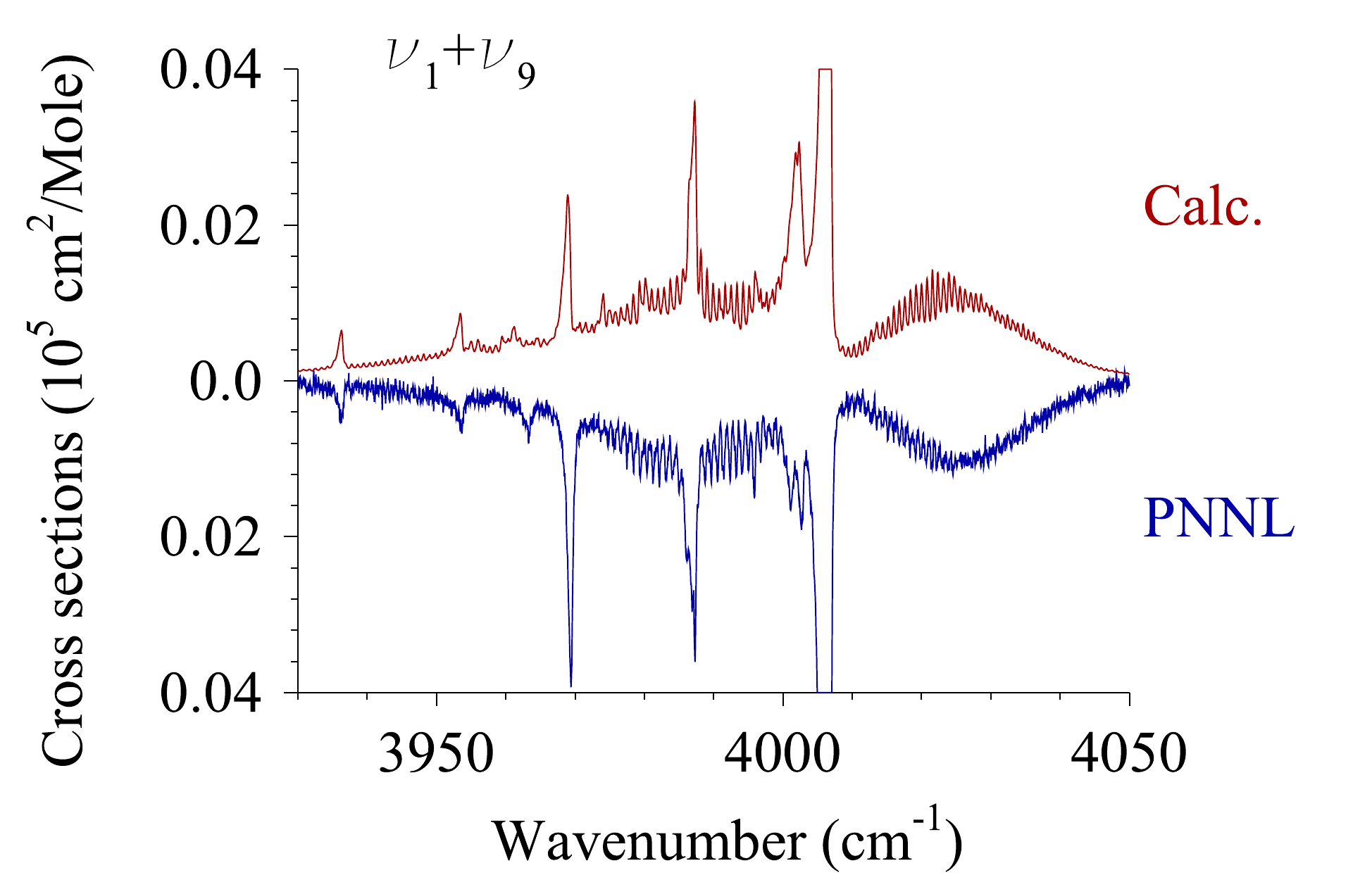}
\caption{Calculated (red curve) and experimental PNNL (blue curve) spectra in
the region of the fundamental, first overtone and lowest combination bands.}
\label{fig:p2}
\end{figure}

Figure~\ref{fig:p2} presents detailed comparisons of our calculated cross
sections   with those of PNNL measured at $T = 298$ K. Our spectra were
converted to cross sections using a Voigt profile  $\sigma = \gamma = 0.075$
cm$^{-1}$ (a half width at half maximum (HWHM) of 0.153 cm$^{-1}$), chosen to
match spectra from the PNNL database \cite{pnnl}. As can be seen, our
calculated spectrum reproduces the PNNL cross sections very well both in the
overall shape and magnitude of the band. This is also true for  the finer
details of the spectrum. For example, the 1800 -- 2000 \cm\ region shows many
features due to hot bands and combination bands which are generally
well-represented in our calculated spectum.

\section{Line list}
\label{s:line-list}

The data necessary to reproduce the spectrum of HNO\3 at temperatures up to
300~K in the 0 - 7000~\cm\ range has been stored in a variety of formats.  In
particular, we have created a line list in the ExoMol format
\cite{jt528,jt548}. The  list contains transitions involving  rotational
quantum number $J$ up to 70 for  $9 \times 10^6$ vibration-rotation energy
levels belonging to 1715 vibrational states and associated transitions
probabilities, in form of Einstein A coefficients. The rotational angular
momentum threshold of $J$ = 70 provides a complete set of rotational energy
levels up to 1050~\cm.  To reduce the very large number of transitions between
different ro-vibrational energy levels we only retain those transitions for
which the intensity is greater than $10^{-32}$~cm/molecule at 296 K. In total
the line list contains about two billion transitions; it can be found on the
ExoMol \cite{jt528} website \url{www.exomol.com}. Key information on the
calculation in form of the initial and refined coefficients of the potential and dipole
moment functions can be found in the supplementary data to this article
\cite{EPAPSHNO3}.

\section{Conclusion}

We present a detailed study of the infrared spectrum of nitric acid.
Calculations are performed using a hybrid variational-perturbation procedure
which allows the whole spectrum can be calculated rapidly on a standard desktop
computer when combined with the initial guess on the intensities of strong
lines provided by experiment. This has allowed us to tune both the potential
energy surface and dipole moment function to the available experimental data.
Comparison with the experimental compilations available in HITRAN \cite{jt557}
and the PNNL database \cite{pnnl} generally give excellent agreement. However
we find that HITRAN is systematically missing features due to hot bands, even
when these are rather strong.

HNO\3 has a strong spectral signature in the Earth's atmosphere which can be
clearly seen from space. As such it is one of a number of species that are
considered to be possible signatures of life (biosignature). To help aid the
detection of life outside the solar system and other studies on hot
astronomical bodies, we are currently preparing an HNO\3 line list which should
be valid over an extended temperature range. This line list will be published
elsewhere \cite{jtHNO3}.

\begin{acknowledgments}
This work was supported by the ERC under Advanced Investigator Project 267219.
\end{acknowledgments}

\clearpage
%\bibliography{journals_phys,jtj,HNO3,exogen,methods,additional,linelists,ap,history,PT,PH3}

\end{document}